\documentclass[12pt,preprint]{aastex}
\usepackage{url}

\def\etal{\hbox{et al.}$\,$}
\def\et{\hbox{et al.}$\,$}

\def\eg{{\it e.g.}}
\def\deg{\ifmmode^\circ\else$^\circ$\fi}

\def\ah{\ifmmode{^\textrm{\scriptsize h}}\else{$^\textrm{\scriptsize
h}$}\fi}
\def\am{\ifmmode{^\textrm{\scriptsize m}}\else{$^\textrm{\scriptsize
m}$}\fi}
\def\as{\ifmmode{^\textrm{\scriptsize s}}\else{$^\textrm{\scriptsize
s}$}\fi}

\newcommand{\Msun}{\ifmmode{\rm M_\odot}\else{M$_\odot$}\fi}
\newcommand{\Zsun}{\ifmmode{\rm Z_\odot}\else{Z$_\odot$}\fi}

\slugcomment{Submitted to The Astronomical Journal}
\shortauthors{Moles et al.}
\shorttitle{The ALHAMBRA Survey}

\begin{document}

\title{The ALHAMBRA Project: A large area multi medium-band\\
optical and NIR photometric survey\footnote{Based on observations
collected at the German-Spanish Astronomical Center, Calar Alto,
jointly operated by the Max-Planck-Institut f\"ur Astronomie
Heidelberg and the Instituto de Astrof\'{\i}sica de Andaluc\'{\i}a
(CSIC).}}


            \author{M. Moles\altaffilmark{1},
	    N. Ben\'{\i}tez\altaffilmark{1,2},
            J.~A.~L. Aguerri\altaffilmark{3},
            E. ~J. Alfaro\altaffilmark{1},
            T. Broadhurst\altaffilmark{4},\\
            J. Cabrera-Ca\~no\altaffilmark{5},            
            F.~J. Castander\altaffilmark{6},
            J. Cepa\altaffilmark{3,7},
            M. Cervi\~no\altaffilmark{1},
            D. Crist\'obal-Hornillos\altaffilmark{1},\\           
            A. Fern\'andez-Soto\altaffilmark{8},
            R.~M. Gonz\'alez Delgado\altaffilmark{1},
            L. Infante \altaffilmark{9},
            I. M\'arquez\altaffilmark{1},
            V.~J. Mart\'{\i}nez\altaffilmark{8,10},\\
            J. Masegosa\altaffilmark{1},
            A. del Olmo\altaffilmark{1},
            J. Perea\altaffilmark{1},
            F. Prada\altaffilmark{1},
            J.~M. Quintana\altaffilmark{1},
            and S.~F. S\'anchez\altaffilmark{11}
            }
\altaffiltext{1}{Instituto de Astrof\'{\i}sica de Andaluc\'{\i}a, 
CSIC, Apdo. 3044, E-18080 Granada}
\altaffiltext{2}{Instituto de Matem\'aticas y F\'\i sica Fundamental, 
CSIC, Serrano 113-bis, Madrid 28006}
\altaffiltext{3}{Instituto de Astrof\'{\i}sica de Canarias, 
La Laguna, Spain}
\altaffiltext{4}{School of Physics and Astronomy, Tel Aviv 
University, Israel}
\altaffiltext{5}{Departamento de F\'{\i}sica At\'omica, Molecular 
y Nuclear, Facultad de F\'{\i}sica, Universidad de Sevilla, Spain}
\altaffiltext{6}{Institut de Ci\`encies de l'Espai, IEEC-CSIC, 
Barcelona, Spain}
\altaffiltext{7}{Departamento de Astrof\'{\i}sica, Facultad de 
F\'{\i}sica, Universidad de la Laguna, Spain}
\altaffiltext{8}{Departament d'Astronom\'{\i}a i Astrof\'{\i}sica, 
Universitat de Val\`encia, Val\`encia, Spain}
\altaffiltext{9}{Departamento de Astronom\'{\i}a, Pontificia 
Universidad Cat\'olica, Santiago, Chile}
\altaffiltext{10}{Observatori Astron\`omic de la Universitat de 
Val\`encia, Val\`encia, Spain}
\altaffiltext{11}{Centro Astron\'omico Hispano-Alem\'an, 
Almer\'{\i}a, Spain}
\email
{moles@iaa.es,benitez@iaa.es,jalfonso@iac.es,emilio@iaa.es,\\
tjb@wise1.tau.ac.il,jcc-famn@us.es,fjc@ieec.fcr.es,jcn@iac.es,mcs@iaa.es,\\
dch@iaa.es,alberto.fernandez@uv.es,rosa@iaa.es,linfante@astro.puc.cl,\\
isabel@iaa.es,vicent.martinez@uv.es,pepa@iaa.es,chony@iaa.es,jaime@iaa.es,\\
fprada@iaa.es,quintana@iaa.es,sanchez@caha.es}

\begin{abstract}
 Here we describe the first results of the ALHAMBRA survey which
 provides {\it cosmic tomography} of the evolution of the contents of
 the Universe over most of Cosmic history. Our novel approach employs
 20 contiguous, equal-width, medium-band filters covering from 3500
 \AA\ to 9700 \AA, plus the standard $JHK_s$ near-infrared bands, to
 observe a total area of 4 square degrees on the sky. The optical
 photometric system has been designed to maximize the number of
 objects with accurate classification by Spectral Energy Distribution
 type and redshift, and to be sensitive to relatively faint emission
 features in the spectrum. The observations are being carried out with
 the Calar Alto 3.5m telescope using the wide field cameras in the
 optical, LAICA, and in the NIR, Omega-2000. The first data confirm
 that we are reaching the expected magnitude limits (for a total of
 100 ksec integration time per pointing) of AB $\leq$ 25 mag (for an
 unresolved object, S/N = 5) in the optical filters from the blue to
 8300~\AA, and from AB = 24.7 to 23.4 for the redder ones. The limit
 in the NIR, for a total of 15 ks exposure time per pointing, is (in
 the Vega system) K$_s$ $\approx$ 20 mag, H $\approx$ 21 mag,
 J$\approx$ 22 mag. Some preliminary results are presented here to
 illustrate the capabilities of the ongoing survey. We expect to
 obtain accurate redshift values, $\Delta z/(1+z) \leq 0.03$ for about
 5 $\times 10^5$ galaxies with I$\leq 25$ (60\% completeness level),
 and $z_{med}$ = 0.74. This accuracy, together with the homogeneity of
 the selection function, will allow for the study of the redshift
 evolution of the large scale structure, the galaxy population and its
 evolution with redshift, the identification of clusters of galaxies,
 and many other studies, without the need for any further
 follow-up. It will also provide targets for detailed studies with
 10m-class telescopes. Given its area, spectral coverage and its
 depth, apart from those main goals, the ALHAMBRA-Survey will also
 produce valuable data for galactic studies.
\end{abstract}


\section{Introduction: Global scientific aim and opportunity}

Only over the last few years it has become possible for Observational
Cosmology to gather enough data on the distant universe to feed our
comprehension of the evolution of the different objects that populate
it. It has become almost commonplace to study protogalaxies at
redshifts $z > 5$, and to observe particular objects at redshifts as
high as $z \approx 6.5$ (Becker \etal 2001, Kashikawa \etal 2006,
Kawai \etal 2006) or even $z \approx 7.5$ (Bradley \etal 2008). At the
same time, samples of objects have been collected through different
techniques at smaller distances (and shorter evolutionary times) from
us, and the different properties of objects in separate redshift
ranges have been measured and compared. However, it remains true that
to this day, no homogeneous sample of objects has been collected
covering a significant range of the age of the universe, even if some
remarkable efforts have been devoted to the production of wide-field,
shallow surveys, that cover the low-redshift end (like 2dFGRS, Colless
\etal 2001, SDSS, York \etal 2000, VVDS, Le F\`evre \etal 2005 or
DEEP-2, Davis \etal 2003) while other groups have directed their
efforts towards the most distant end, through very deep, small-area
surveys like the HST Deep Fields or other legacy programs (Ferguson
\etal 2000, Beckwith \etal 2006).

The Cosmological Principle implies the existence of maximally
symmetric subspaces and the existence of a one-to-one relation between
redshift and time. The corresponding evolutionary nature of the
depicted Universe is a model-independent prediction, prior to any
consideration about the value of the cosmological parameters. A direct
way to tackle many of the problems posed by modern cosmology is hence
to materialize a {\sl foliation of the space-time}, producing narrow
slices in the $z$-direction whereas the spatial sections are large
enough to be cosmologically representative, obtaining as output a kind
of {\em Cosmic Tomography}.

From the observational point of view, to trace {\em Cosmic Evolution},
which is a central topic in Cosmology, the genuine evolutionary
effects have to be disentangled from both the physical variance at a
given redshift and the details of the metric as measured in --- or,
depending on the point of view, induced by --- the cosmological
model. In other words, to approach the question of evolution
meaningfully it is necessary to sample large volumes even at low
redshift, to capture not only representative average properties but
also their variance. This will necessarily imply a survey featuring a
combination of wide area and depth, and a continuous spectral coverage
to avoid complex selection functions that depend on the redshift and
on the nature of the objects under analysis. Moreover, the quest for
the necessary precision implies high enough spectral resolution and
photometric accuracy.

Up to now, the largest surveys ensuring complete spectral coverage for
large samples have been photometric, and done with broad-band
filters. The resulting redshift precision obtained with these
techniques ($\sim 0.03$ in $\Delta z/(1+z)$, at best, see Cucciati
\etal 2006 and Ilbert \etal 2006) and in Spectral Energy Distribution
(SED) determination are correspondingly rough. Moreover, large area
photometric surveys like SDSS are necessarily shallow, whereas deeper
surveys have sampled the distant and/or faint Universe in rather small
areas. At the other extreme in spectral resolution, spectroscopic
surveys can neither go as deep as the photometric ones nor cover large
enough areas. Moreover, they are defined in order to observe a
restricted spectral region, producing a selection effect that is a
function of the object type and redshift that can be very intricate
due to the selection effects inherent to spectroscopy
(Fern\'andez-Soto \etal 2001).

For those scientific purposes where the detailed properties of
individual objects are not the goal, the aim from an observational
stand is therefore that of finding the optimal filter combination to
produce the most homogeneous, deepest, and most accurate possible
photometric survey. Such a survey would produce precise enough values
for the redshift and SED for large numbers of objects. We present here
the \textbf{A}dvanced \textbf{L}arge \textbf{H}omogeneous
\textbf{A}rea \textbf{M}edium \textbf{B}and \textbf{R}edshift
\textbf{A}stronomical, \textbf{ALHAMBRA}-Survey, that intends to
produce such an optimum survey for the study of cosmic evolution. It
has been designed to achieve (with the facilities at hand) the best
compromise between large area and depth, good spectral resolution and
coverage, in order to produce an optimum output in terms of redshift
and SED accuracy. The ALHAMBRA-Survey is a deep photometric survey
using 20 contiguous, equal-width, medium-band optical filters from
3500 \AA\ to 9700 \AA, plus the three standard broad band ($JHK_s$)
NIR filters. The total area surveyed by ALHAMBRA will be 4 square
degrees, being therefore placed halfway in between traditional imaging
and spectroscopic surveys.

By design, the ALHAMBRA-Survey will provide precise ($\Delta z <
0.03(1+z)$) photometric redshifts and SED classification for several
hundred thousand galaxies and AGNs, allowing for different kinds of
analysis regarding populations and structures, and their evolution in
time.  The details of the project, including simulations and expected
results, and all the related aspects are described in the
ALHAMBRA-Book that can be found at \url{http://www.iaa.es/alhambra}. Thanks
to the unbiased nature of this survey (i.e. neither designed to detect
a given class of objects nor to be precise only in some fixed spectral
window), important problems other than cosmic evolution can be
addressed. These include the study of stellar populations in the
galactic halo, the search for peculiar stellar objects, ranging from
very cold stars to blue stragglers, and the possible detection of
debris from galactic satellites in the Milky Way halo. Moreover, the
large surveyed volume and the ability to finely discriminate between
different spectral energy distributions will permit the serendipitous
detection of objects that could be classified as {\em exotic} or {\em
  rare}. This broad category includes very high redshift galaxies
($\approx$ 2500 objects at $z > 5$, with $\Delta z < 0.01$, expected
from scaled HDF observations) and QSOs.

The observations are being carried out with the 3.5m telescope of the
Centro Astron\'omico Hispano-Alem\'an, CAHA, Calar Alto Observatory
(Almer\'{\i}a, Spain) and the wide-field imagers in the optical
(LAICA) and in the NIR (Omega-2000). The collected data render
possible the study of many different astronomical problems in a
self-contained way and will provide with very interesting targets for
individual studies with large size telescopes.

A separate article (Ben\'{\i}tez \etal 2008) deals with the selection
of the optical filters and the optimization of their characteristics
to maximize the spectral information, while in this work we present
the main characteristics that specifically define the ALHAMBRA-Survey,
including some preliminary results from the data we have already
accumulated.

This paper is organized as follows: in Section 2 we present the
project implementation and its present status, and in Section 3 the
first, preliminary results. We compare the ALHAMBRA survey with other
surveys in Section 4, whereas our conclusions are presented in Section
5.
%
%

\section{Survey design, implementation and status}

\subsection{Description of the filter system}

The idea to use photometric information to determine the redshift of
faint sources was first proposed by Baum (1962), and later re-launched
by Loh \& Spillar (1986), Koo (1986) and Connolly \etal (1995) as a
{\sl poor person} redshift machine. Its importance and adequacy to
produce relevant data to different cosmological analysis was
recognized after, among others, the works by Lanzetta \etal (1996),
Connolly \etal (1997), or Fern\'andez-Soto \etal (1999), on the Hubble
Deep Field. Nowadays, many if not most of the surveys which have
already been completed or are under development, include the use of
broad-band filters and photometric redshift techniques.

Hickson \etal (1994) were the first to discuss the possibility of
using a set of medium-band filters to continuously cover a large
spectral range, and produce photometric data that could be considered
equivalent to a very-low resolution spectrum for each detected
object. No discussion was done however in that work on the number and
kind of filters that would be needed in order to optimize the output
in terms of z and SED accuracy for a given instrumental setup and
observing time. Later the surveys CADIS (Meisenheimer \et 1998) and
COMBO-17 (Wolf \et 2001a, Wolf \et 2001b, Wolf \et 2004, Bell \et
2004) used a combination of broad- and medium-band filters with
similar purposes. In particular, COMBO-17 employs the standard broad
band filters ($UBVRI$) plus 12 narrow- or medium-band filters sampling
several spectral domains between 4000 and 9200~\AA.  At the end, in
this and similar surveys the full spectral coverage is obtained via
broad band filters. COMBO-17 has reached very good quality in getting
photometric redshifts with an accuracy of 1\% in $\Delta z/(1+z)$ at
$R<21$. It covers an area of 1 square degree. The degradation in
quality at fainter magnitudes is planned to be compensated with the
use of 4 redder filters ($Y, J1, J2$ and $H$) in the MANOS-deep survey
(or COMBO-17+4) covering a field of 0.77 square degrees.

The ALHAMBRA-Survey was designed to cover all the visible spectral
domain with equally wide, contiguous medium-band filters to optimize
its scientific output in terms of accuracy of the z and SED
determinations. By design, it is possible to detect, other than the
overall SED, relatively faint emission lines. As already pointed out,
the details are explained in a separated paper by Ben\'{\i}tez \et
(2008), where different implementations were devised and analyzed to
get the best solution for a given fixed total amount of observing
time. The resulting optical filter system includes a total of 20
contiguous, medium-band, FWHM = 310~\AA, top-hat filters, that cover
the complete optical range from 3500 to 9700 ~\AA. The optical
coverage is supplemented with the standard NIR $JHK_{s}$ filters.

\begin{figure*}
\centering
\includegraphics[angle=-90,width=14cm]{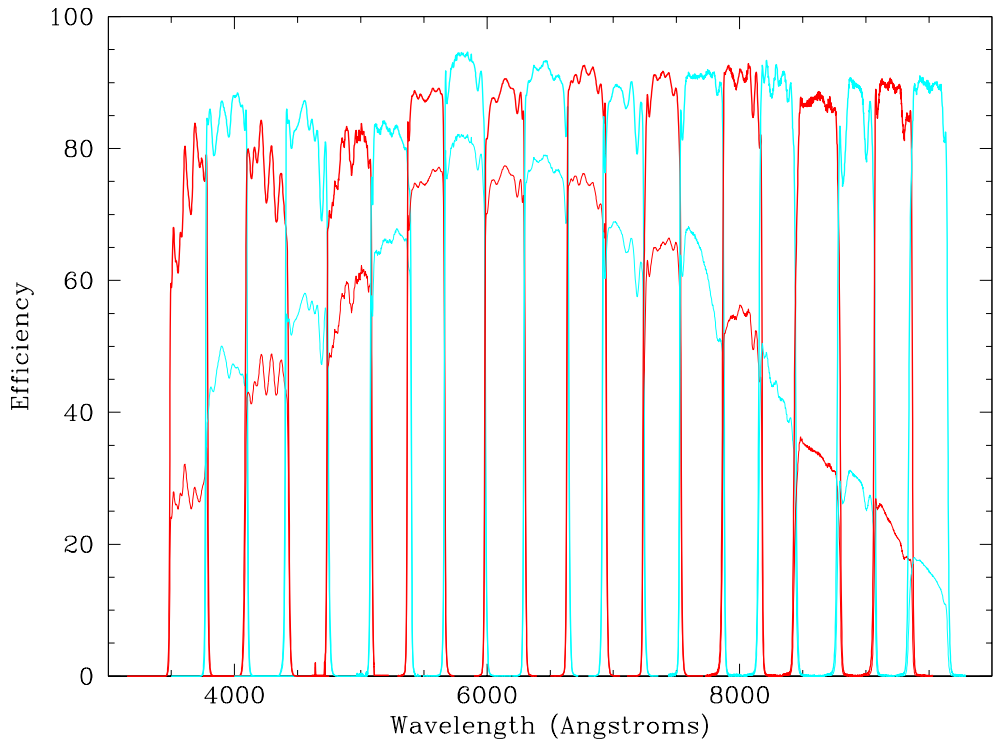}
\caption{Transmission curves for one of the optical filter sets for
the ALHAMBRA-Survey as measured in the laboratory. We also show the
effective total transmission (lower curve), after taking into account
the quantum efficiency of the CCD detector, the atmosphere transmission
(at Air Mass = 1.3) and the reflectivity of the primary mirror of the
Calar Alto 3.5m telescope}
\label{filters}
\end{figure*}

The filter set was designed having in mind a series of restrictive
requirements about their spectral shape, homogeneity, and
transmission. In particular, all filters should have very steep side
slopes, close to zero overlap in $\lambda$, a flat top, and
transmissions in excess of 70\%. Given the particular disposition of
the four detectors in the LAICA focal plane, four sets of filters had
to be produced, in such a way that all of them were equivalent within
strict limits. The complete set of $20 \times 4$ filters was
manufactured by BARR Associates.  They were confirmed to be within
specifications at the laboratory of the Instituto de Astrof\'{\i}sica
de Andaluc\'{\i}a. The transmission curves are shown in
Figure~\ref{filters}.

\subsection{Expected redshift precision and the NIR filters}

\begin{figure*}
\centering
\includegraphics[angle=-90,width=14cm]{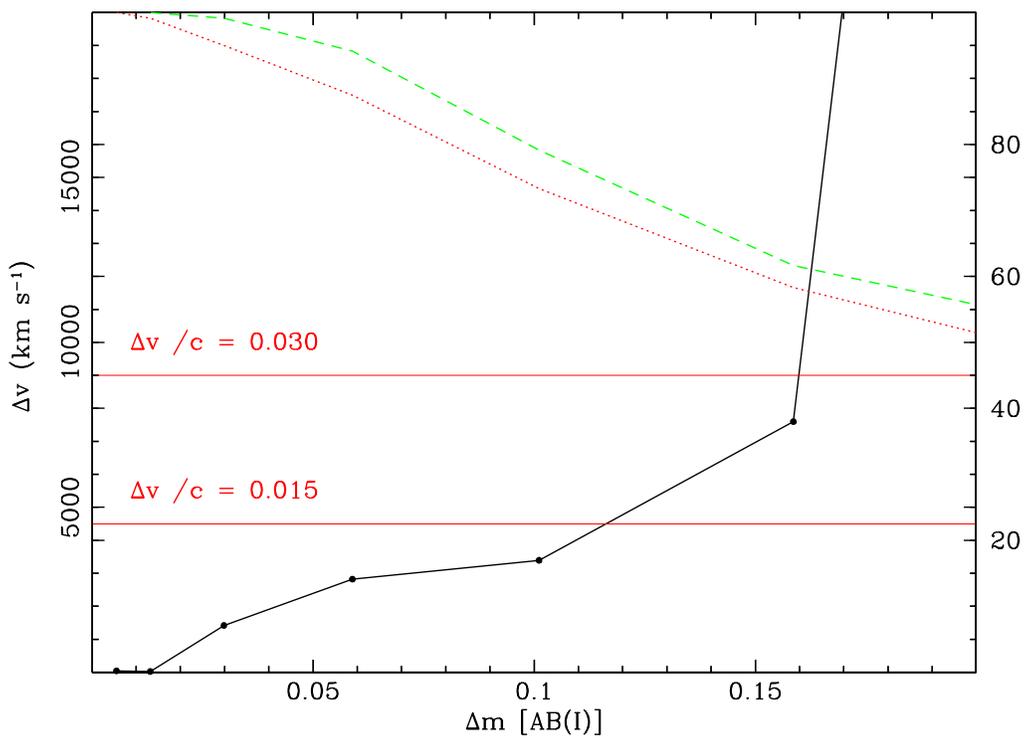}
\caption{Estimates of the photometric redshift dispersion as a
  function of the photometric uncertainty in the $I$ band. The black
  dots and continous line show the velocity uncertainty, while the red
  dotted (green dashed) line shows the percentage of objects with a
  given $I$-band uncertainty that have redshift residuals smaller than
  $\Delta z/(1+z)= 0.015$ (0.030).}
\label{errors}
\end{figure*}

For a survey with the characteristics of ALHAMBRA the quality of the
final results depends critically on the photometric errors and the
adequacy of the templates used to compute photometric redshifts. To
test the first aspect we have created a grid of galaxy spectra using
the templates presented in Ben\'{\i}tez (2000), with redshifts between
0 and 5.0, and absolute $K_s$-band magnitudes ranging from -23 to
-16. All these galaxies have been "observed" through the full ALHAMBRA
photometric system, with noise added to the observed fluxes according
to the properties of our data. Each galaxy had then its ALHAMBRA
photometric redshift measured via the same code used by Fern\'andez-Soto
\etal (1999), and the offset between the original and the calculated
values of the redshift was obtained. We have then binned the objects
according to their photometric $I$-band magnitude uncertainty, and
estimated the dispersion around the correct redshift value for each
bin. The results are presented in Figure~\ref{errors}, where it can be
seen that we obtain an excellent redshift accuracy ($\Delta z/(1+z) <
0.03$) even for objects with magnitude errors as large as $\Delta
AB(I) \approx 0.15$---the error we measure for typical objects at
AB($I$)$\approx 24.5$. Moreover, the percentage of all those objects
that actually have redshift residuals smaller than that is larger than
50\%--recall that the redshift error distribution is strongly
non-Gaussian, with long tails and second peaks due to the so-called
"catastrophic errors". This analysis, of course, does not include the
possible systematic effects induced by the choice of the template set.

\begin{figure*}
\centering
\includegraphics[angle=0,width=8cm]{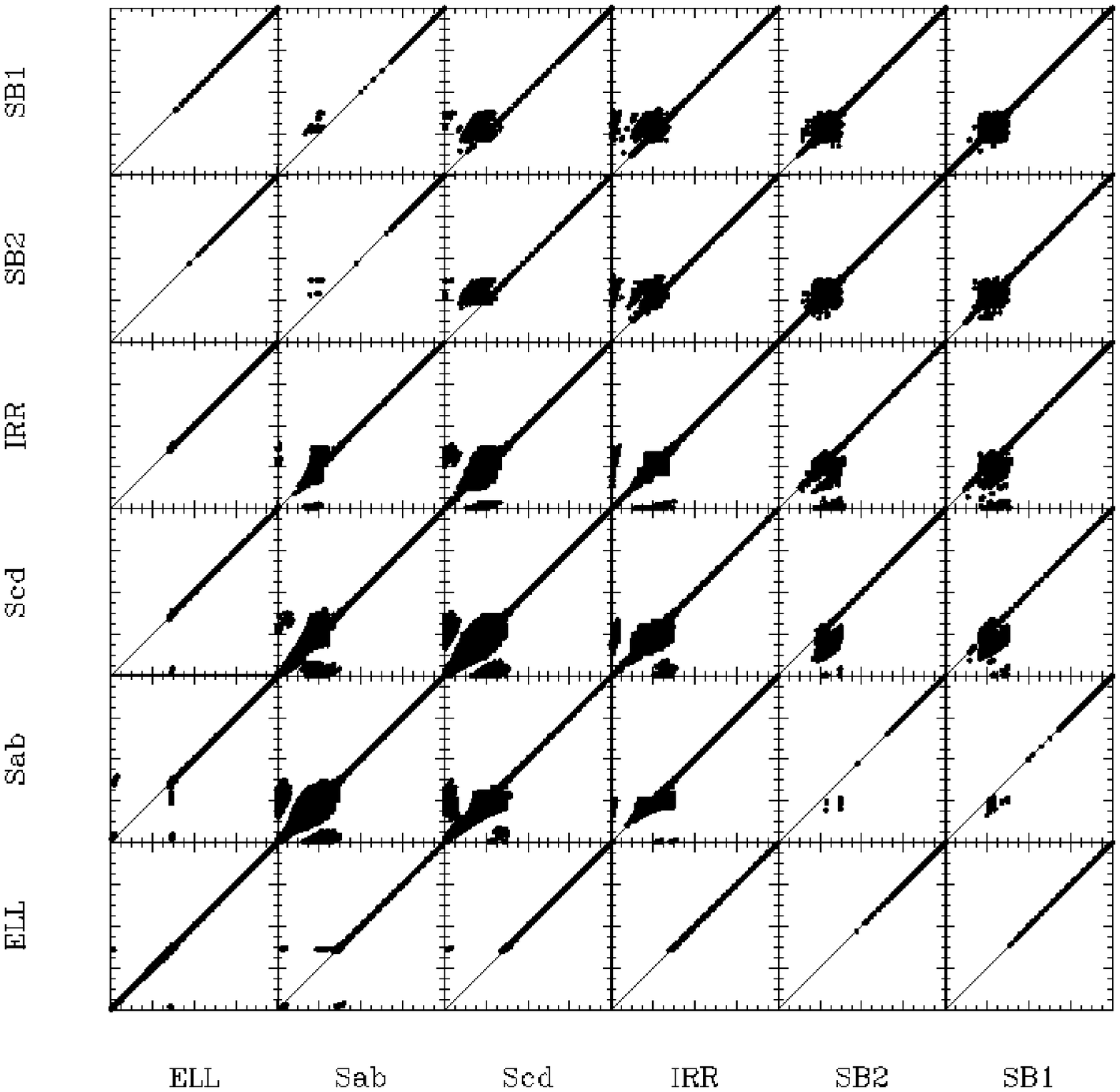}
\includegraphics[angle=0,width=8cm]{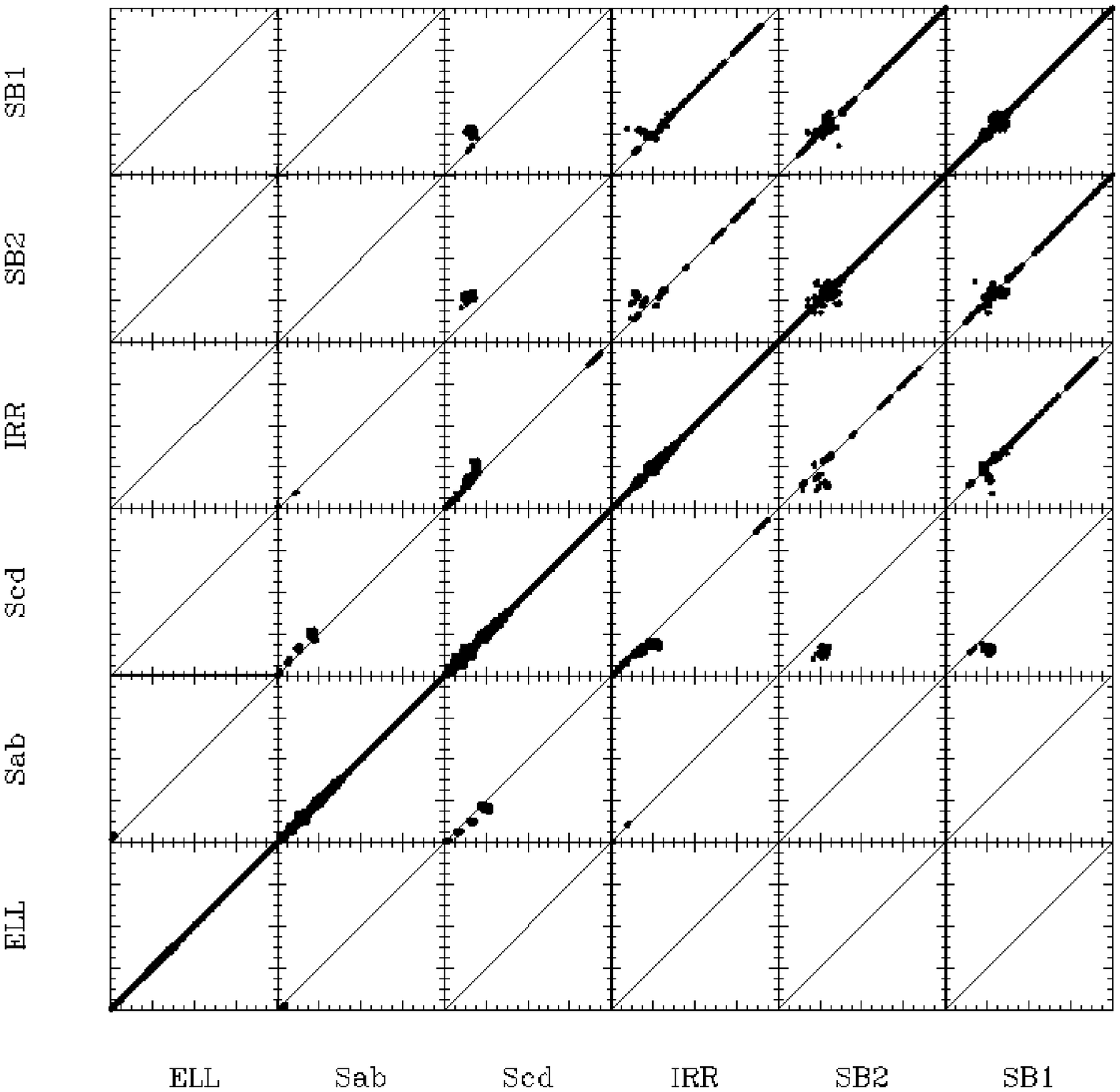}
\caption{Theoretical degeneracies in type and redshift expected for
galaxies in the ALHAMBRA survey measured to an accuracy of 0.2
magnitudes in all filters. The upper panel shows the case where no
infrared information is available, and the lower panel corresponds to
the case where the NIR information is included. Each of the $6 \times
6$ sub-panels corresponds to a $z_1$ (photometric redshifts) vs $z_2$
(actual values) diagram with redshifts ranging from 0 to 8.}
\label{infrared}
\end{figure*}

Inclusion of the NIR information can significantly improve the
determination of the redshift in some cases. In particular, as pointed
out by many authors using photometric redshift techniques in deep
surveys (\eg Coe \etal 2006), the use of NIR filters can help to break
the degeneracy between low-redshift ($z \approx 0.5$) and
high-redshift ($z \approx 3$) galaxies. The reason behind this
degeneracy is the possible confusion between the Balmer and Lyman
breaks, which are the most salient features of the respective spectral
energy distributions. In absence of any infrared information, it is
not possible to determine the slope of the rest-frame red end of the
spectrum, the range that can in fact tell the difference between both
families. Each one of the left-side panels in Figure~\ref{infrared}
explicitly shows the degeneracies between the different spectral types
and redshifts. For each of the six galaxy types used in the exercise
(rows and cells) and every redshift (from $z=0$ to $z=8$ in both axes
in every box), the points mark the types that degenerate with it in
color space---for this particular figure, a magnitude ``thickness''
$\Delta m=0.2$ has been chosen, so that any two galaxy spectra that
differ in less that $\Delta m$ in all the ALHAMBRA filters considered
have been taken as degenerate, and hence indistinguishable.

It can be seen in the right-side panels how (for the $\Delta m=0.2$
case) the presence of NIR data eliminates most of the degeneracies
between low and high redshift, and sharply separates the elliptical,
Sab, and Scd galaxies from the rest and from each other, leaving only
some residual degeneracy between the three bluest types. Of course,
the infrared information also adds greatly to the scientific content
of the survey, via the more direct relation existing between the
galactic mass and the infrared luminosity.

Our preliminary results (see next section) show us that the
simulations above should not be far off the mark; our $I$-band
photometric errors are $<10\%$ up to $I\approx$ 24, with expected
redshift errors $\delta z/(1+z) \leq $1.5\%. We also expect that most
of the objects up to I$\leq$25 will have photometric redshift errors
$<3\%$ .


\subsection{The Survey area}

It is well known that astronomical objects cluster on the sky on
different scales. The clustering signature contains a wealth of
information about the structure formation process. A survey designed
to cope with the cosmic variance and to describe and understand the
clustering needs to probe as many scales as possible, up to the
homogeneity scale. So, on one hand, searching contiguous areas is
important in order to cover smoothly the smallest scales where the
signal is stronger and to obtain an optimally-shaped window
function. But, on the other hand, measuring a population of a certain
volume density is a Poissonian process with an associated variance
and, in principle, one would obtain different densities of the same
population when measuring at different places. The variance in those
measurements is dictated by the volume density of the population under
study, the volume searched and the clustering of the population. In
order to beat down this cosmic (or, technically, sampling) variance
one needs to survey independent volumes. So, all in all, a balance
must exist between probing contiguous and independent areas.

For a survey aiming at resolving a number N$_{st}$ of SED-types, the
sampling errors in any counting statistical measure will go as

$$\sigma_{s} = K \sqrt{\frac{N_{st}}{N_f\;\Delta z}}$$

\noindent where $N_f$ is the number of fields of a given area and
$\Delta z$ the redshift interval to be resolved. To get a quantitative
evaluation of this statistical noise, we have used the SDSS local
luminosity function, that yields, with a plausible parameterization of
the evolution, a redshift distribution similar to that obtained from
the HDF at our magnitude limit. Expressing N$_{f}$ as the number of 1
square degree fields, we find K in the range 0.01 - 0.02 for redshift
between 0.3 and 2. Therefore, if we intend to have relative sampling
errors no larger than a few percent at any redshift, with a redshift
resolution better than $\Delta z = 0.05$, N$_{f}$ should be similar to
N$_{st}$, which results in a total field of a few square degrees.
%
\begin{table}[tb]
\caption{The ALHAMBRA-Survey Fields. We include together with the name 
a reference to overlaps with other surveys of interest.}
\centering
\begin{tabular}{lcccccc}
\hline
Field name  &  RA(J2000) & DEC(J2000)& 100 $\mu$m & E(B-V) & l & b \\
\hline
ALHAMBRA-1  & 00 29 46.0 & +05 25 28 & 0.83 & 0.017 &  113 & -57 \\
ALHAMBRA-2/DEEP2  & 02 28 32.0 & +00 47 00 & 1.48 & 0.031 & 166 & -53  \\
ALHAMBRA-3/SDSS  & 09 16 20.0 & +46 02 20 & 0.67 & 0.015 & 174 & +44 \\
ALHAMBRA-4/COSMOS  & 10 00 28.6 & +02 12 21 & 0.91 & 0.018 & 236 & +42 \\
ALHAMBRA-5/HDF-N   & 12 35 00.0 & +61 57 00 & 0.63 & 0.011 &  125 & +55 \\
ALHAMBRA-6/GROTH   & 14 16 38.0 & +52 25 05 & 0.49 & 0.007 &  95 & +60 \\
ALHAMBRA-7/ELAIS-N1 & 16 12 10.0 & +54 30 00 & 0.45 & 0.005 & 84 & +45 \\
ALHAMBRA-8/SDSS & 23 45 50.0 & +15 34 50 & 1.18 & 0.027 & 99 & -44 \\
\hline
\end{tabular}
\end{table}
%
%
%

%
\setcounter{footnote}{1}

 On the technical side we had to consider the geometry of the imager
LAICA, with four 4k$\times$4k CCDs, arranged in a $2 \times 2$ mosaic
with the gap between two adjacent (vertical and horizontal) CCDs being
almost the same size of the CCD itself. Thus, one pointing corresponds
to 4 patches of 15.4' $\times$ 15.4' over a total field of 44.4'
$\times$ 44.4', such that with four pointings a total, contiguous area
of 1 degree squared is covered\footnote
{\url{http://www.caha.es/CAHA/Instruments/LAICA/index.html}}. This geometry
imposes a minimum contiguous area patch of 1 degree $\times$ 0.25
degrees and produces 2 such strips in two pointings. The Omega-2000
camera used for NIR
observations\footnote{\url{http://www.mpia-hd.mpg.de/IRCAM/O2000/index.html}}
covers a field of view equivalent to one of the LAICA CCDs.

Considering the sampling constrains together with the technical
characteristics of the detectors, the expected efficiency of the
atmosphere-telescope-imager, and the amount of available time for the
project, we finally decided to cover 2 such strips in each of the 8
selected fields, a total of 4 square degrees to ensure a large enough
area coverage and good sampling.

The fields were selected taking as first and basic criterion their low
extinction, as measured in Schlegel \etal (1998). Then, within the
lowest extinction patches, we tried to identify those containing
neither bright sources nor conspicuous structures. We decided that a
significant overlap with other surveys would be an asset, to ensure
the possible cross-checking of our results and maximum
complementarity, in particular regarding the largest coverage in
wavelength.  The selected fields are listed in Table 1. The 100 $\mu$m
emission and E(B-V) values quoted in the table are the median for the
whole 1$^{\square}$ field centered at the given position.  Only
ALHAMBRA-1 is new in the sense of having no overlap with other
surveys.
 

At the moment of writing this article (Feb 2008) we have already obtained about
64\% of the near infrared data and about 35\% of the visible data. The
NIR observations with Omega-2000 were started in August 2004, whereas
we could start taking visible data with LAICA only in September
2005. We expect to complete our observations by 2010.

We will discuss in Section 3 below some preliminary results obtained
with our first complete $15'\times 15'$ pointing, including
observations with all the ALHAMBRA filters.

\subsection{The calibration strategy}

As mentioned above, we have already completed the first LAICA pointing
in all the filters. In the quoted ALHAMBRA-book we have presented the
simulations we performed to design the survey and to find the expected
depth of the data for the strategy chosen. We are now in the position
of checking the quality of the data delivered by the survey, the
validity of our simulations and the adequateness of our observational
strategy

Indeed, a crucial aspect to do that analysis is the photometric
calibration of the data. This is a particularly demanding aspect of
the project. Apart from other considerations, it has been shown that
the reliability of the photometric redshifts depends critically on the
photometric accuracy (Coe \etal 2006). This aspect is particularly acute
in our case since we are in fact introducing a new photometric system
in the optical domain. Thus, special care is needed to anchor it to
existing photometric systems and to primary calibrators.

To define the reference fluxes and magnitudes of the ALHAMBRA
photometric system we have chosen a set of primary stars from the
lists by Oke \& Gunn (1983), Oke (1990), Massey \& Gronwall (1990) and
Stone (1996), together with the 4 fundamental calibrators adopted by
the HST. The list includes the standard star BD+17$^o$4708, the
primary calibrator of the SDSS system.

We give here a short account of calibration procedure, whereas the
details will be presented in a forthcoming article. The devised
procedure starts with the selection of stars in our frames that have
SDSS photometry. These will play the role of secondary standards. We
plan to obtain spectrophotometric observations of all these stars,
calibrated with respect to those primary standards. The spectra will
then be fitted by stellar models and the ALHAMBRA-system colors will
be obtained from the fitted models by integration over the filters. We
will also compare the synthetic u'g'r'i'z' colors with the SDSS values
to check the consistency between the systems. In this way we expect to
have a calibration accurate to the 2\% level or better. The comparison
between photometric and spectroscopic redshifts for the galaxies with
spectroscopic observations will also be used as a check of the
calibration of the ALHAMBRA data (Coe \etal  2006).

We will set the ALHAMBRA magnitudes on the AB system (Oke \& Gunn
1983),
$$AB_{\nu}=-2.5\log f_{\nu}-48.60$$
\noindent
where $f_{\nu}$ is the flux per unit frequency from an object in erg
s$^{-1}$ cm$^{-2}$ Hz$^{-1}$.

The magnitudes will be defined with reference to their
spectrophotometric data by adopting
$$ m = -2.5 \rm{log}
\frac{\int_{F}f(\nu)S_F(\nu)d(log\nu)}
     {\int_{F}S_F(\nu)d(log\nu)}             + Cte$$
\noindent
where S$_F(\nu)$ is the transmission curve corresponding to the 
atmosphere-telescope-filter-detector combination.

This is in fact the usual way to calibrate narrow-band images when the
photometric system is not previously defined (see for example,
M\'arquez \& Moles 1996). The SDSS Consortium has also adopted this
strategy to define their own photometric system (Fukugita \et 1996;
Smith \et 2002).

At the moment of writing the present paper we have not yet completed
the observations of secondary stars to get the final calibration. In
order to get a preliminary calibration that would be robust and
accurate enough to allow the analysis of the main aspects of the
survey in terms of depth and reliability, we have implemented a
different procedure. Briefly, what we have done is to select 228 flux
calibrated stellar spectra from the Hubble's Next Generation Spectral
Library (NGSL; Gregg \etal 2004) covering a wide range of physical
properties. From these spectra synthetic colors were obtained in
both, SDSS and ALHAMBRA systems and the first set of transformation
equations between the two systems were obtained. The results are given
in Table~\ref{calibr}.

Then we have applied these equations to a set of stars with accurate
SDSS and ALHAMBRA instrumental magnitudes to obtain the zero point of
the photometric calibration. The zero points found in this way are
accurate at the level between 5 and 10\%.

\begin{table}
\small
\caption{The calibration relations. The numbers are the coefficients
of the relation between any given ALHAMBRA magnitude and the SDSS
values (blanks for null coefficients)}
\vskip 1truecm
\centering
\label{calibr}
\begin{tabular}{r r r r r r r}
\hline
Filter
 & \multicolumn{1}{c}{Orig.}
 & \multicolumn{1}{c}{u}
 & \multicolumn{1}{c}{g}
 & \multicolumn{1}{c}{r}
 & \multicolumn{1}{c}{i}
 & \multicolumn{1}{c}{z}
\\
\hline
 01 &  -0.0208 &   0.9314 & \null & \null &   0.0690 & \null \\
 02 &   0.0257 &   0.3626 &   1.3321 &  -0.7066 & \null & \null \\
 03 &  -0.0467 &   0.1442 &   1.3099 &  -0.4560 & \null & \null \\
 04 &  -0.0391 & \null &   1.0607 & \null & \null &  -0.0587 \\
 05 &   0.0147 & \null &   0.6488 &   0.5480 & \null &  -0.1961 \\
 06 &  -0.0336 & \null &   0.6341 &   0.3667 & \null & \null \\
 07 &  -0.0192 & \null &   0.3199 &   0.7865 &  -0.1056 & \null \\
 08 &  -0.0057 & \null &   0.0604 &   1.1792 &  -0.2391 & \null \\
 09 &  -0.0050 & \null & \null &   1.1009 &  -0.1008 & \null \\
 10 &   0.0067 & \null & \null &   0.7279 &   0.2720 & \null \\
 11 &   0.0188 & \null &  -0.2000 &   1.1309 &   0.0682 & \null \\
 12 &   0.0156 & \null &  -0.0876 &   0.6046 &   0.4811 & \null \\
 13 &  -0.0284 & \null &   0.0839 &  -0.1523 &   1.0693 & \null \\
 14 &   0.0208 &  -0.0178 & \null & \null &   1.0177 & \null \\
 15 &   0.0247 & \null & \null &  -0.2473 &   1.2463 & \null \\
 16 &   0.0372 & \null & \null &  -0.3725 &   1.3717 & \null \\
 17 &   0.0161 & \null & \null & \null &   0.1777 &   0.8207 \\
 18 &  -0.0528 & \null & \null & \null &  -0.0383 &   1.0413 \\
 19 &  -0.0199 & \null & \null & \null &  -0.1200 &   1.1207 \\
 20 &   0.0186 & \null & \null & \null &  -0.2436 &   1.2433 \\
\hline
\end{tabular}
\end{table}
\normalsize

The NIR data have been calibrated through the 2MASS catalog stars
present in our frames. The rms of the calibration is always below
0.04 mag in all 3 bands.

%


\subsection{Observing Strategy}

\begin{figure*}
\centering 
\includegraphics[angle=-90,width=14cm]{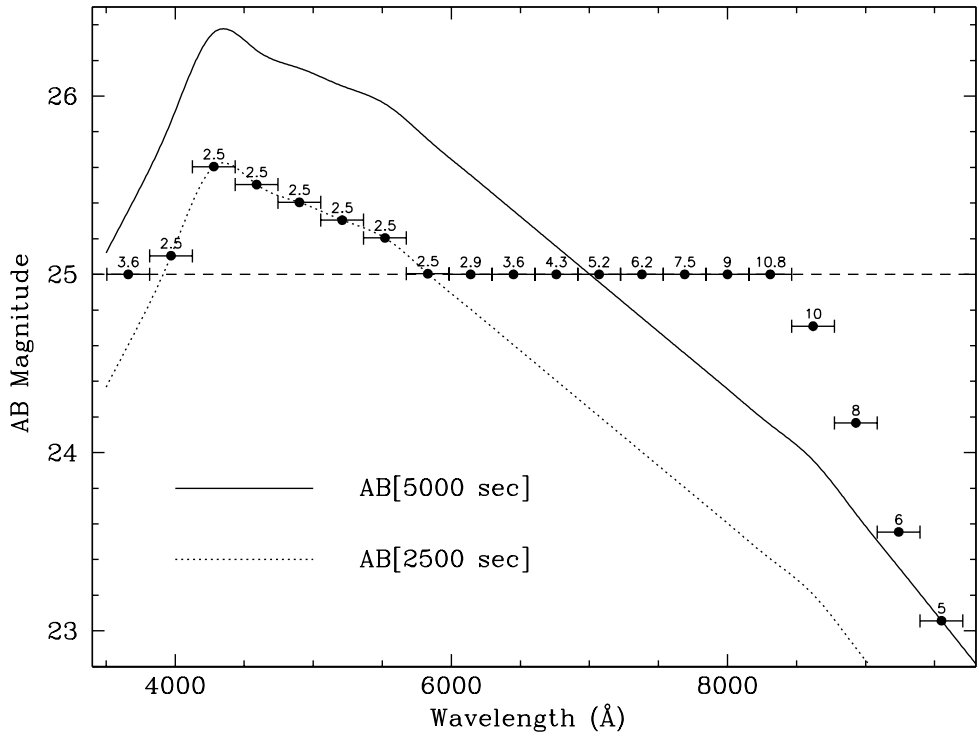} 
\caption{Expected limit magnitudes for the ALHAMBRA filters, at S/N =
5, estimated for our survey. The exposure time per filter (in ksec)
is given for every filter.}
\label{exptimes}
\end{figure*}

As was mentioned in Section 2.1, the ALHAMBRA filter set was devised
following detailed simulations in order to optimise the quality and
number of photometric redshifts for galaxies in our fields. Similarly,
the exposure times per filter were defined in such a way that the
expected magnitude limits in each one would be as homogeneous as
practically possible. This driving principle, combined with a minimum
exposure time of 2.5 ksec per filter (due to the observing logistics),
and the known fact that our system is less efficient in the red end of
the spectrum, propped us to divide the total LAICA time per pointing
(100 ksec) in the way presented in Figure~\ref{exptimes}. Thus, the
target $AB=25$ limit would be reached or even exceeded for all filters
bluewards of $\approx 8500$ \AA, and redwards from there the
decreasing efficiency makes the limit magnitude drop until AB$\approx
23$ at $\approx 9500$ \AA. In the near infrared, previous experience
with Omega-2000 decided us to divide the available 15ksec per pointing
equally into the $J, H,$ \and $K_s$ filters, aiming to reach $J=22,
H=21, K_s=20$ (in the Vega-based system). These expectations will be
compared with the real data in Section 3.

Our image quality limit to consider the data as "scientifically valid"
is, in real time at the telescope dome, a value of the seeing which
should be $<1.4"$. However, the seeing, as well as the transparency,
is measured {\it a posteriori} during the pipeline reduction, and the
individual images are kept or eliminated from the final combination
based on a more complex quality assessment, that will be presented
elsewhere.

\subsection{Data reduction}

For the reduction of the images two different pipelines have been
used. In the case of the LAICA data, the preliminary standard steps of
overscan correction and flat-fielding are performed. After that, the
illumination correction using smooth superflats is applied. In the
optical filters redder that 7000 \AA\ the images suffer from a
substantial fringing pattern that is removed using the procedure
described in Alcala \etal (2002).  This fringing pattern contributes
less than 2\% to the flat-fields, so we have not corrected the
flat-fields at this stage.

In the case of the NIR filters we first removed the dark current
frames and divided by a normalized superflat constructed combining the
science images.  An additive pupil ghost is present in the superflat
images, that is more prominent in the $J$ band ($\sim 5\%$
contribution). The pupil ghost is fitted in the normalized flat-field
images using the mscpupil task in the IRAF MSCRED package (Valdes
2002). The adjusted pupil pattern ($p_{ij}$) is removed from the
flat-field dividing it by $(1+p_{ij})$, and is subtracted from the
individual images after scaling it by the median background
level. After having flat-fielded the images, the sky structure of each
individual image is removed with the XDIMSUM package (Stanford,
Eisenhardt \& Dickinson 1995) using the sky image constructed with the
median of the 6 closest images, that in case of the $J,H,K_s$ filters
correspond to timescales of 480,360,276 seconds respectively. Before
removing the background, the sky image is median filtered using
$5\times5$ window in order to reduce its noise. During this process
also masks for each individual image with the location of the cosmic
rays are created. The cosmic rays and bad pixels are fixed to proceed
with the next steps.

At this stage, the sky level is measured, and SExtractor (Bertin \&
Arnouts 1996) is used to obtain an initial estimate of the FWHM of
each image, and the number of sources above a given S/N. The relative
transparency of each image is measured using high S/N stars. These
numbers are used to remove bad images or images out of the survey
requirements. After that, the astrometry of a reference image is
calibrated using the USNO-B1.0 (Monet \etal 2003) catalog. The
external astrometric rms is $\sim 0.12$ arcsecs in each axis. The
remaining images are calibrated internally, in a first iteration using
the reference image in the filter. After completing an ALHAMBRA
pointing a deep image constructed using images with good FWHM and
transparency in several selected filters will be created and used
afterwards to calibrate the internal astrometric solution. For this
purpose we have used our own algorithm to match the image sources with
those in the external catalog, and the IRAF ccmap task to obtain an
order 3 polynomial solution. The internal astrometric solution rms is
$\sim 0.03$ arcsec in each axis for the Laica images and $\sim 0.05$
arcsecs in the Omega-2000 ones. Finally, the astrometry of each
individual image is re-calibrated internally with the reference image
using Scamp\footnote{\url{http://terapix.iap.fr}} obtaining similar
results. For image combination
Swarp\footnote{\url{http://terapix.iap.fr}} is used. This software
takes into account the distortion pattern present in the wcs headers,
and allows the user to obtain resampled images with a different pixel
size, in the desired sky projection.  Also the previously computed
relative transparency is used to uniform the zero points of the
individual images. In our process, before performing the final
combination using Swarp, the resampled images are used to apply a
pixel rejection algorithm in other to improve the cosmic ray rejection
and bad pixel masks. In the case of the Omega-2000 images, with a
roughly double pixel size, we also require a final image with the
Laica pixel scale. The final images for each ALHAMBRA pointing in the
23 filters are registered within the internal astrometric solution
rms.

\section{Survey performance: preliminary results}

The ALHAMBRA survey and filter system design was preceded by
realistic, thorough simulations presented in Ben\'\i tez \etal 2008
and in the ALHAMBRA-book.  Recently we completed the observations of
the first pointing in the 20+3 filters, comprising 4 times a 15'4x15'4
field, and we are thus able to compare the performace of the survey
with our expectations.

\begin{figure*}
\centering
\includegraphics[angle=0,width=14cm]{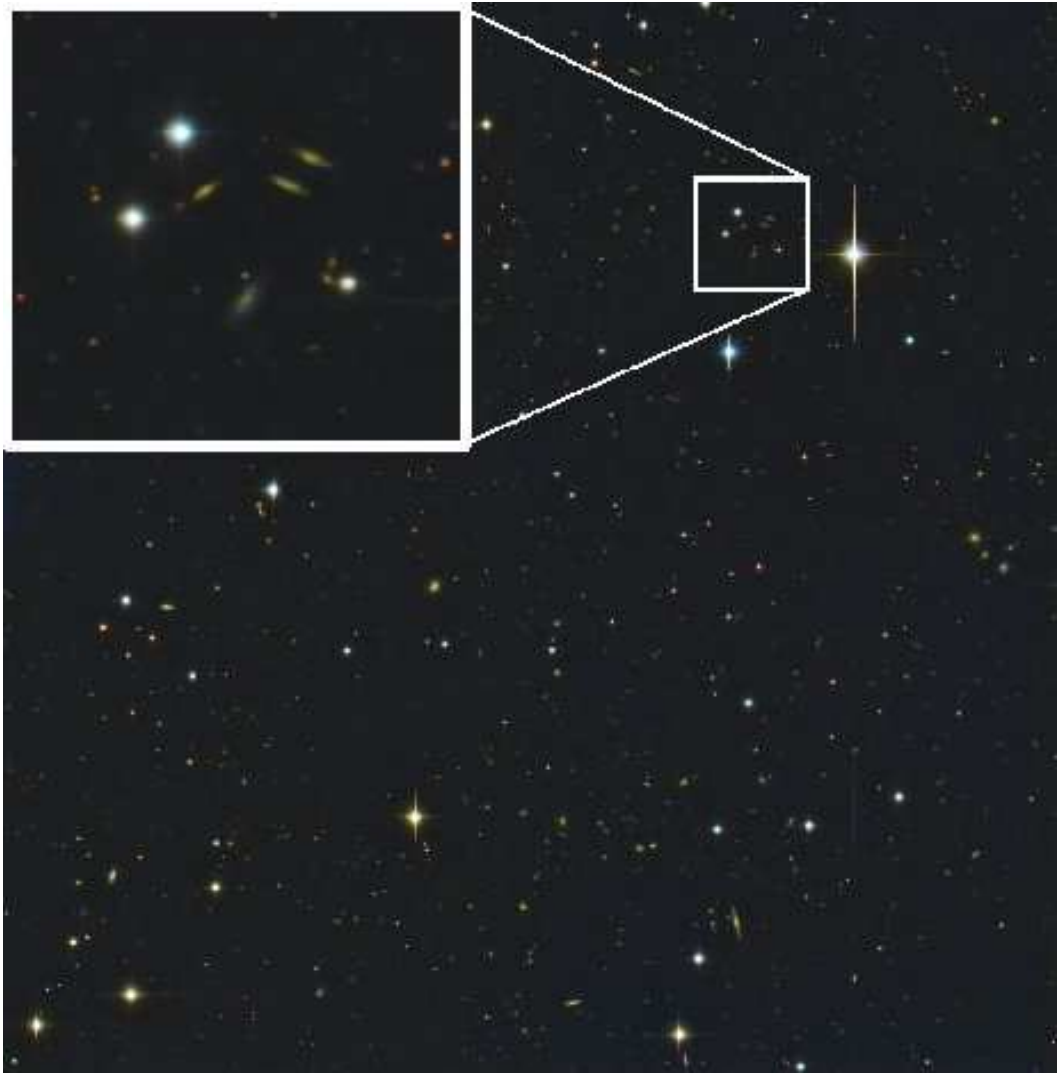}
\caption{The first complete pointing of the ALHAMBRA-8 field in a
square region of 15 arc-minutes side. This color image has been created
making use of data from 14 out of the 23 filters. In the insert,
corresponding to a region of about 1.5 arc-minutes, a small group of
galaxies can be seen}
\label{field}
\end{figure*}

We show in Figure~\ref{field} a color image corresponding to one of
the CCDs of the first complete pointing in the ALHAMBRA-8 field (see
Table 1). An enlargement of this image of approximately 1.5
arc-minutes side, where a small group of spiral galaxies can be seen,
is shown as an insert in the same figure.

These data allow us to go from simulations to actual measurements, and
describe several key characteristics of the survey; others will have to
be complemented with simulations until we carry out follow-up
observations, primarily due to the paucity of available spectroscopic
redshifts in this particular field.

\subsection{Photometric depth}

The data reduction procedure involves the standard steps, including accurate flatfielding and 
defringing (for the red filters) and precise astrometry. SExtractor, with
just standard settings for the different parameters, was used to get
the photometry of the detected objects that we discuss here. The
pipelines and procedures will be described in a separate paper, and
will be made publicly available. Then the data were photometrically
calibrated following the procedure sketched in the previous section.

The ALHAMBRA filter system was primarily designed to obtain the best
determination of $z$ and SED for a fixed amount of observing time per
pointing (100 ksec). We estimated the expected limit magnitudes taking
into account the average extinction in Calar Alto, the performance of
the 3.5m telescope-LAICA system for airmass = 1.3, a final image
quality of FWHM $\approx 1^{\prime\prime}.2$ and the measured
filter$+$CCDs transmission curves. The goal was to get homogeneous
magnitude limits for as many filters as possible, with the restriction
that the minimum exposure time per filter should not be less than 2.5
ksec. We expected to reach AB $\le$ 25 (S/N = 5, point-like source)
for 16 filters ranging from 3500 \AA~ to 8500 \AA.  In the case of the
four reddest filters (close in wavelength to the $z$ band) we could
obtain decreasingly lower limit magnitudes, down to AB = 23.4 in the
last filter centered at 9550 \AA~ (see the quoted ALHAMBRA-book and
Moles \etal 2005).

\begin{figure*}
\centering
\includegraphics[angle=-90,width=14cm]{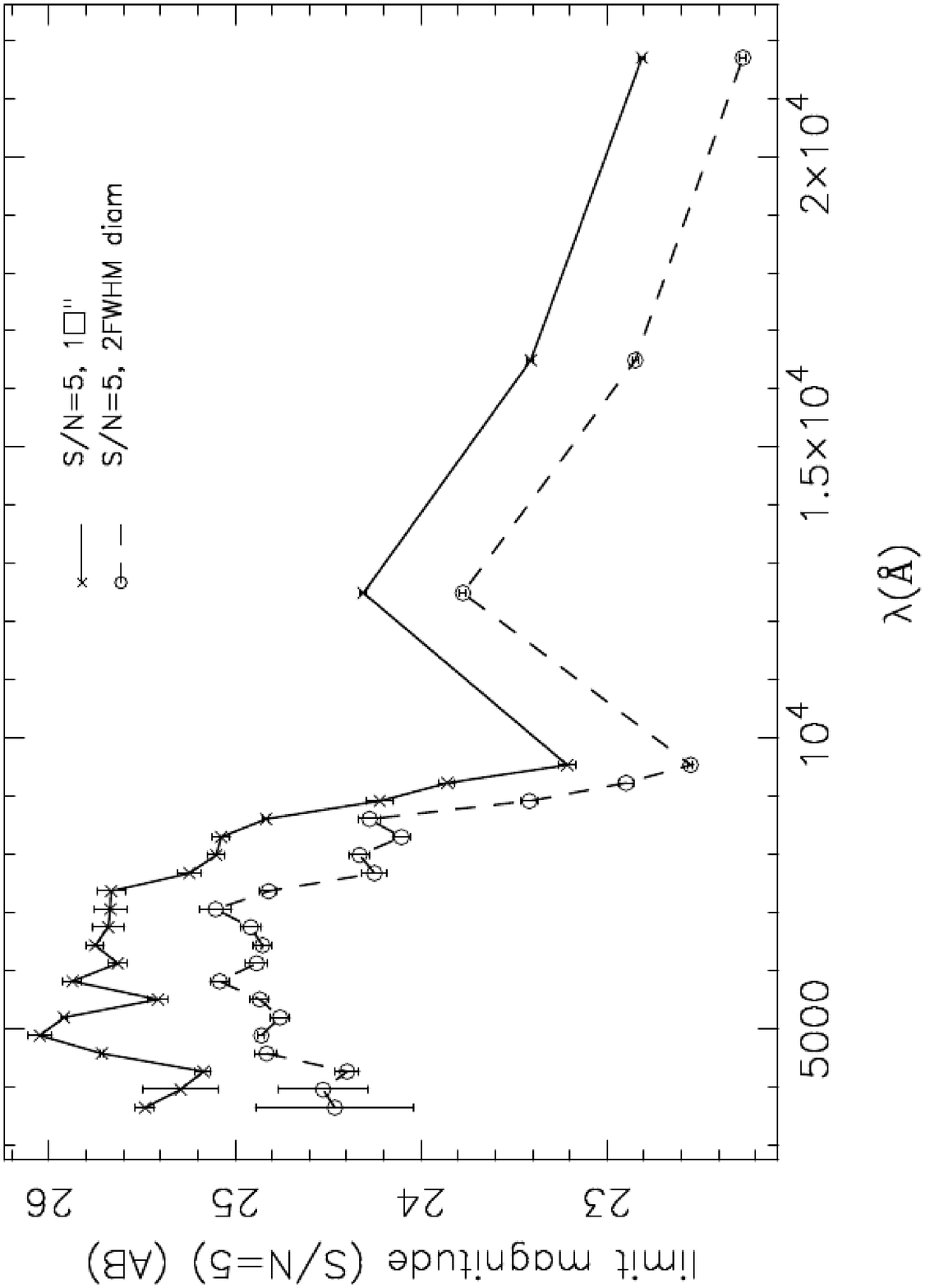}
\caption{The limit magnitude, at S/N = 5 in each filter, for 1" square
aperture and for an aperture of 2$\times$FWHM diameter. The total
observing time per pointing was fixed to 100 ksec for the 20 optical
filters, and to 15 ksec for the NIR bands}
\label{limit}
\end{figure*}

 To ascertain in a quantitative way the actual depth we reach in
each of the filters we have defined the limiting magnitude as that
corresponding to the rms within a 1 square arc-second aperture, at the
S/N =5 level. The values we find are plotted in Figure~\ref{limit}. As
can be seen, we are actually reaching the expected limits in all the
filters, including the reddest ones. We also show in the same figure
the magnitude limit values for a 2$\times$FWHM aperture, which may
represent a more realistic integration area for typical flux measurements.


\begin{figure*}
\centering
\includegraphics[angle=-90,width=14cm]{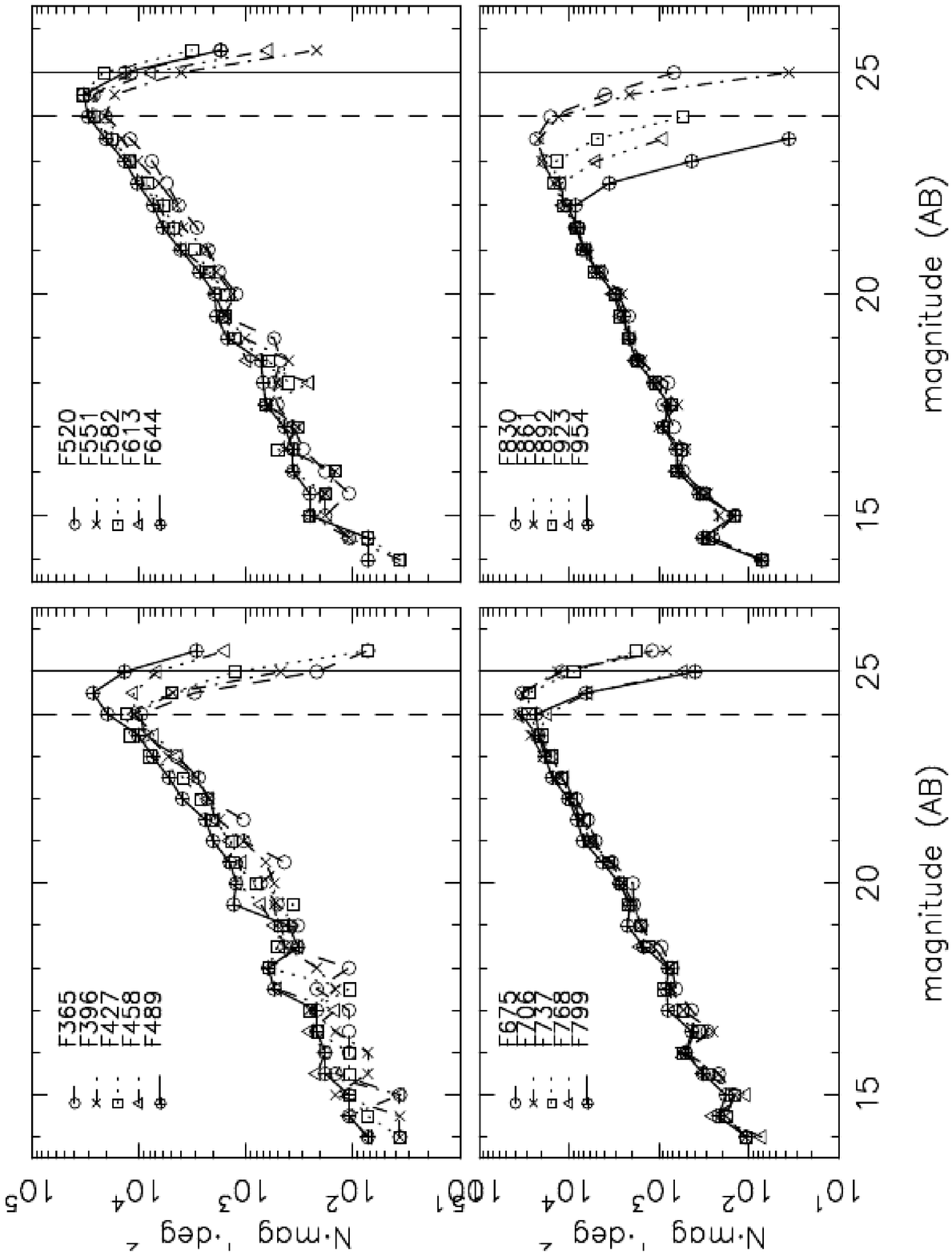}
\caption{Histograms of number of sources to S/N=5 per square degree
per magnitude for the 20 ALHAMBRA optical filters.}
\label{histo}
\end{figure*}

 A complementary way to illustrate the depth of the data is to
consider the number counts in each band. In Figure~\ref{histo} we show
the preliminary number count histograms in the 20 ALHAMBRA optical
filters. The counts have been normalized to unit magnitude and unit
area (square degree). No corrections due to completeness or spurious
detection have been applied.  The histograms indicate that we are
detecting sources at AB $\leq$ 25 in all the filters from the bluest
to about 8000\AA. The limiting magnitude then becomes brighter, and we
end up with AB$\sim$23 at 9500\AA.

\begin{figure*}
\centering
\includegraphics[angle=0,width=14cm]{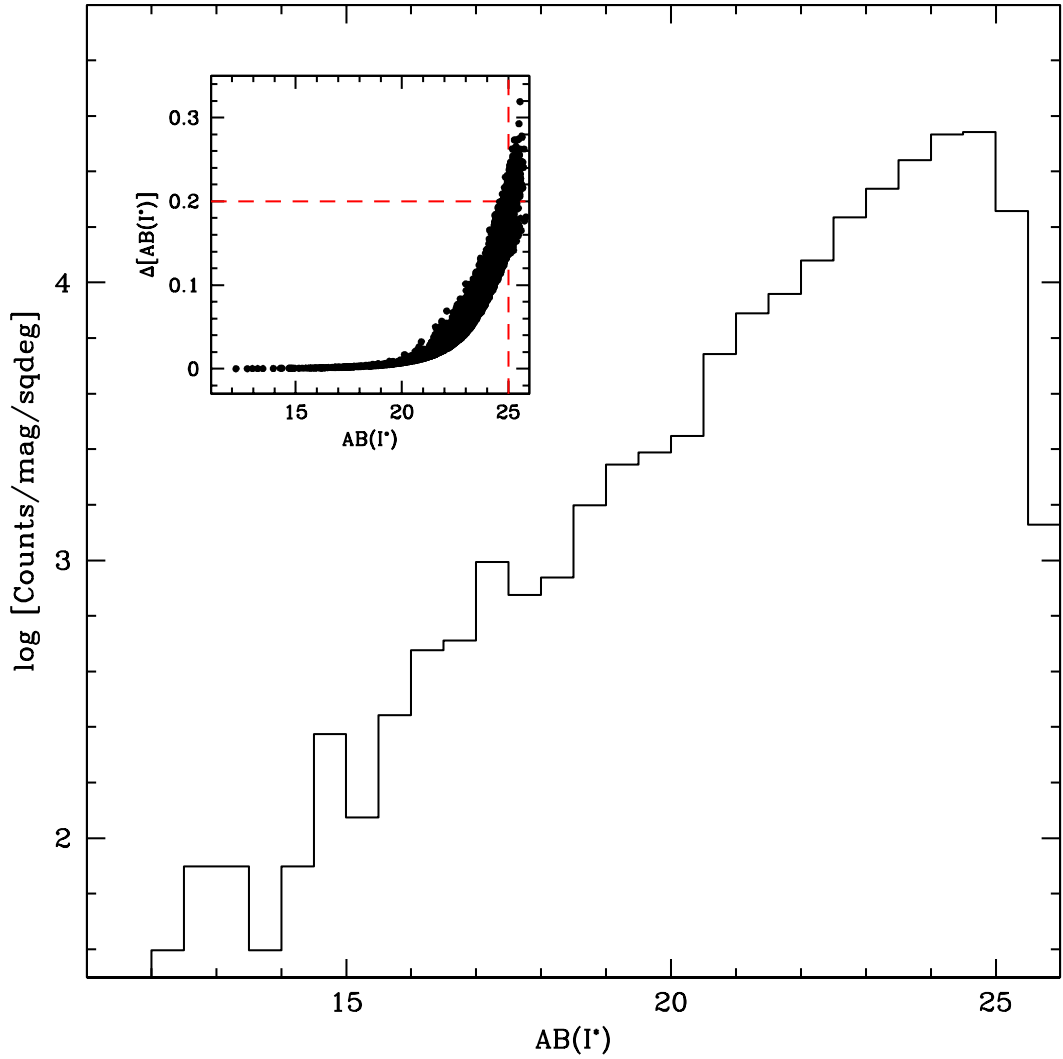}
\caption{Counts as a function of isophotal magnitude in the synthetic
  I-band image. The completeness limit is close to I(AB) = 25. In the
  inserted panel the photometric errors as a function of the isophotal
  I magnitude are shown.}
\label{I-band}
\end{figure*}

Moreover, to be able to compare with broad-band and spectroscopic
surveys (see later), we have synthesized broad-band filters from our
medium-band ALHAMBRA filters. In Figure~\ref{I-band} we present the
results for the $I^*$ (synthetic $I$) band. It can be seen that the
survey is complete to $I^* \approx$ 25 with errors $\approx20$\% in
the worst cases.

\begin{figure*}
\centering
\includegraphics[angle=-90,width=14cm]{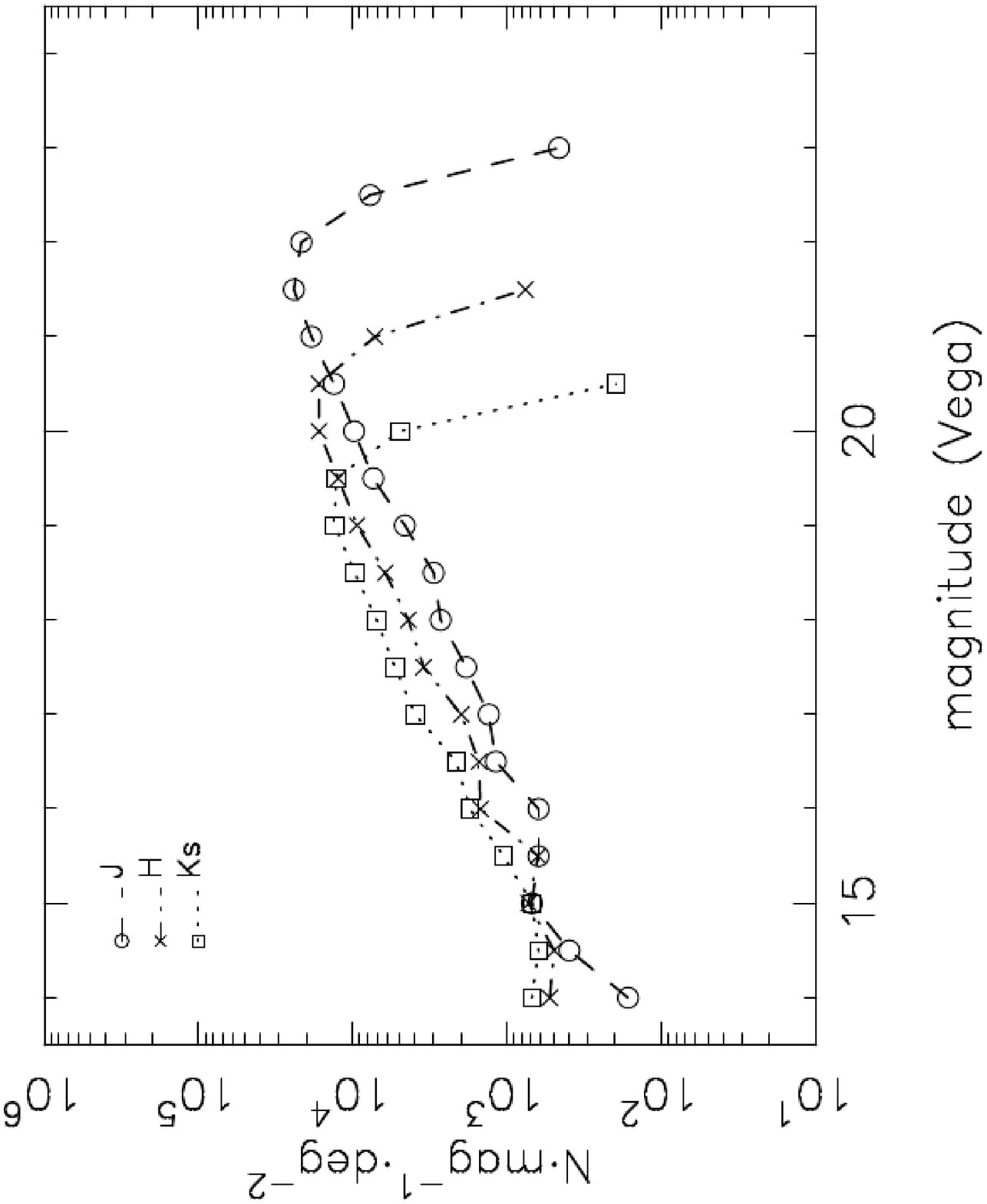}
\caption{Histograms of number of sources to S/N=5 per square degree
per magnitude in the three NIR filters.}
\label{NIR}
\end{figure*}

In the NIR, based on existing experience with Omega-2000, we fixed the
total exposure time to 5 ksec per filter, in order to reach K$_s$
$\approx$ 20, H $\approx$ 21, J $\approx$ 22 in the Vega system (S/N =
5, point like source). The limit magnitudes presented in
Figure~\ref{limit} and the histograms presented in Figure~\ref{NIR}
show that these limits have been reached, and even
exceeded\footnote{Recall that the magnitude difference (AB-Vega) for
the $J$, $H$, and $K_s$ filters is $\approx$ 0.9, 1.5, and 2.0
respectively.}. A more detailed analysis of the NIR galaxy counts will
be presented in Crist\'obal-Hornillos \etal (2008, {\it in
preparation}).

%

\subsection{Photometric redshift depth and accuracy}

\begin{figure*}
\centering
\includegraphics[angle=0,width=12cm]{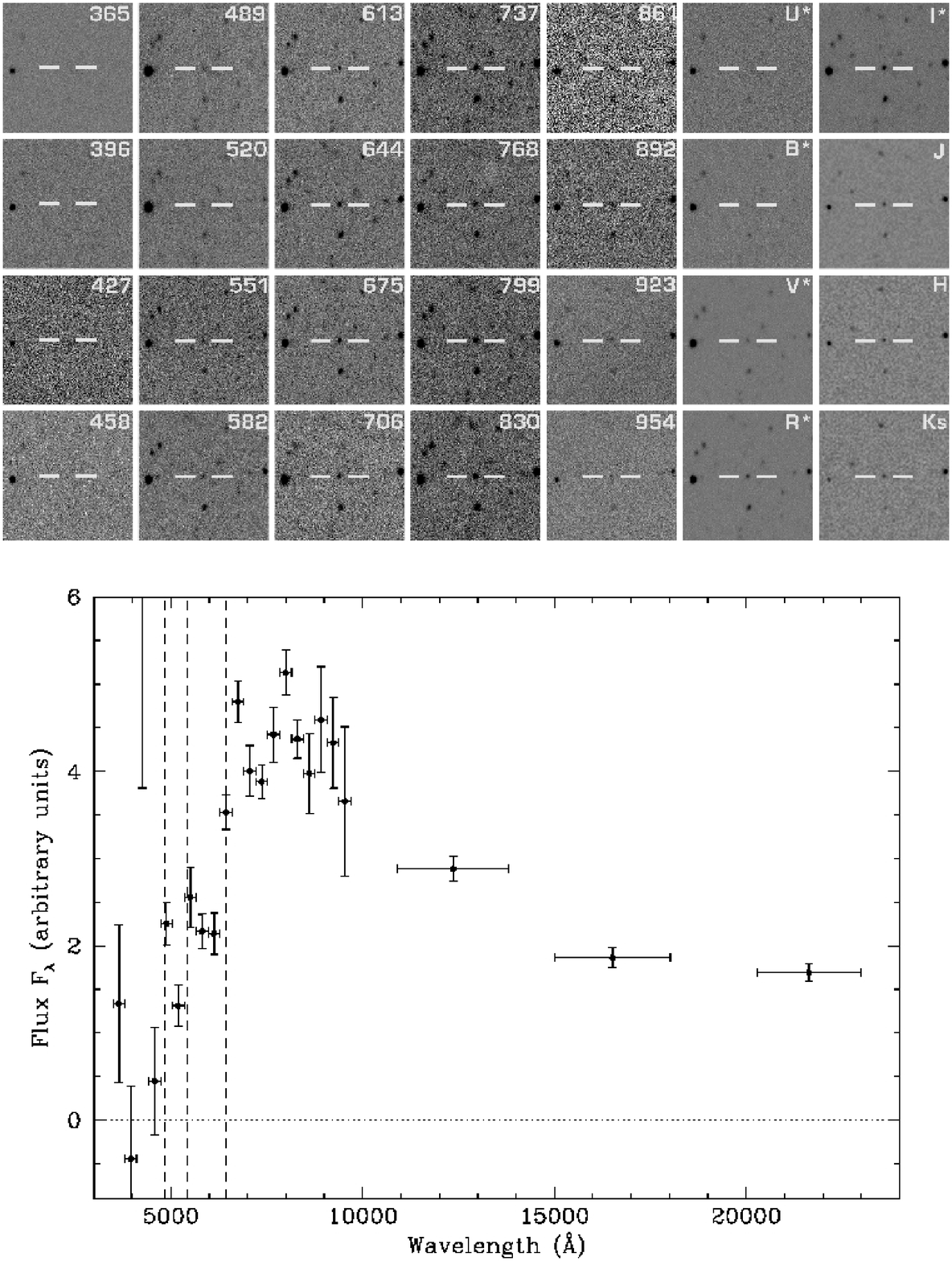}
\caption{Upper panels: Images through all 23 ALHAMBRA filters and the
  synthetic Johnson filters for a single object in the ALHAMBRA-8
  field, marked with horizontal ticks. Each 0.7 arcmin square
  thumbnail is labeled with its wavelength in nanometers, or the
  filter name. Lower panel: ALHAMBRA-spectrum of the same object. The
  vertical dashed lines mark the putative positions of Lyman-$\alpha$,
  Lyman-$\beta$, and the Lyman limit, at the measured redshift
  $z=4.32$. Note that the 427nm image and data point are noisier
  because we did not use the complete exposure time for this filter in
  this preliminary reduction.}
\label{obj03858}
\end{figure*}

The central goal of the ALHAMBRA-survey is to measure with precision
the observed photometry of as many objects as possible; this enables us
to estimate accurate redshifts and spectral types.  The use of a large
number of filters, contiguous and with minimal overlap among
themselves provides a clear, inequivocal representation of the galaxy
SED.  This is well illustrated by Figure~\ref{obj03858} which shows a
small (0.7 $\times$ 0.7 arc-minutes) thumbnail image of a
$AB(I)\approx23$ object in the ALHAMBRA-8 field in all 23 ALHAMBRA
images and the synthetic $U^*B^*V^*R^*I^*$ filters, together with a
calibrated ALHAMBRA ``spectrum'' of the object, a starburst with
$z=4.23$.

The customary way of determining the photo-z accuracy of a survey is
comparing estimates with a large sample of spectroscopic redshifts;
unfortunately, the first field which we have completed only has SDSS
spectroscopy available, too shallow and sparse for that purpose, and
we plan to carry out spectroscopic follow-up observations for this and
other fields.

\begin{figure*}
\centering
\includegraphics[angle=-90,width=14cm]{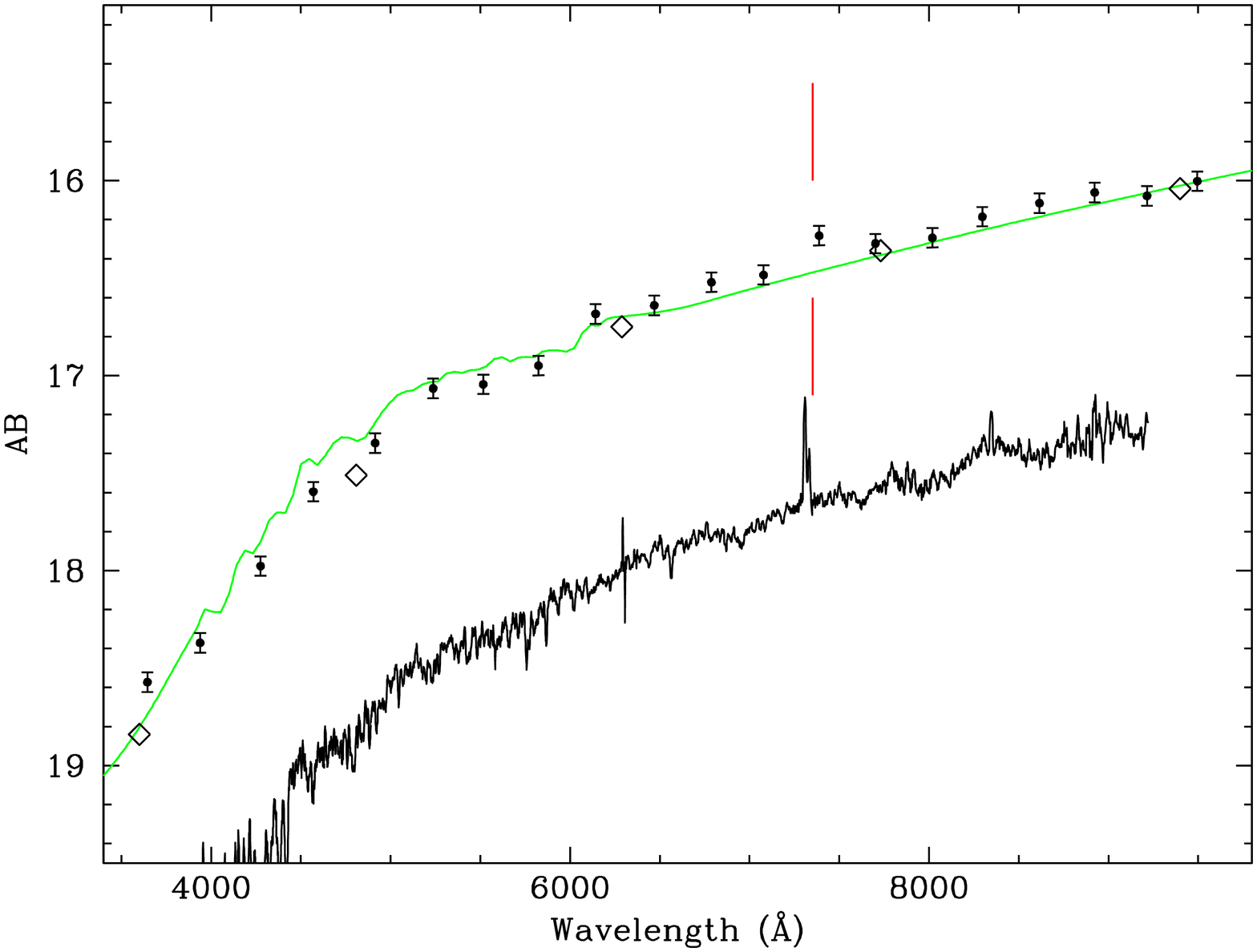}
\caption{ALHAMBRA-spectrum (points with error bars), best-fit model
  (smooth line); and SDSS photometry (diamonds) and spectrum (lower,
  noisy line) for one of the galaxies with spectroscopic redshift in
  the ALHAMBRA-8 field. The vertical ticks mark the expected position
  of the H$\alpha$+N{\sc II} complex at the best-fit redshift. Notice
  the apparent excess in the ALHAMBRA photometry at the expected
  wavelength, even though the total equivalent width measured for the
  complex in the SDSS spectrum is only 15 \AA.}
\label{spec}
\end{figure*}

The NASA Extragalactic Database contains nine galaxies with
spectroscopic redshifts within the limits of our $15'\times 15'$
pointing, all of them SDSS galaxies (Adelman-McCarthy \etal 2007),
with redshifts in the range $z \lesssim 0.2 $. Despite the inadequacy
of this sample for statistical purposes it allows us to give the
reader a taste of the redshift accuracy of the ALHAMBRA survey (see
Figures ~\ref{zphot} and \ref{spec}).

We are presently using three different codes to measure photometric
redshifts, based on different methods, in order to eventually choose
the most robust and accurate strategy.

a) One is the code developed by Fern\'andez-Soto \etal (1999), based
on galaxy template fitting with an extended Coleman, Wu and Weedman
(1980) library comprising six templates.

b) The second one is the BPZ program, as described in Ben\'{\i}tez
(2000), which uses a Bayesian magnitude-type prior and the empirically
calibrated template library described in Ben\'{\i}tez \etal (2004)
and complemented with two very blue starburt types (Coe \etal 2006),
eight galaxy templates in total.
 
c) The third method has been recently implemented by S\'anchez (2008,
{\it in preparation}), and uses a synthetic library which also
includes stellar spectra, AGN types and dust reddening. The use of a
synthetic library provides information about the metallicity, the star
formation history, and other characteristics of the object.

\begin{figure*}
\centering
\includegraphics[angle=0,width=14cm]{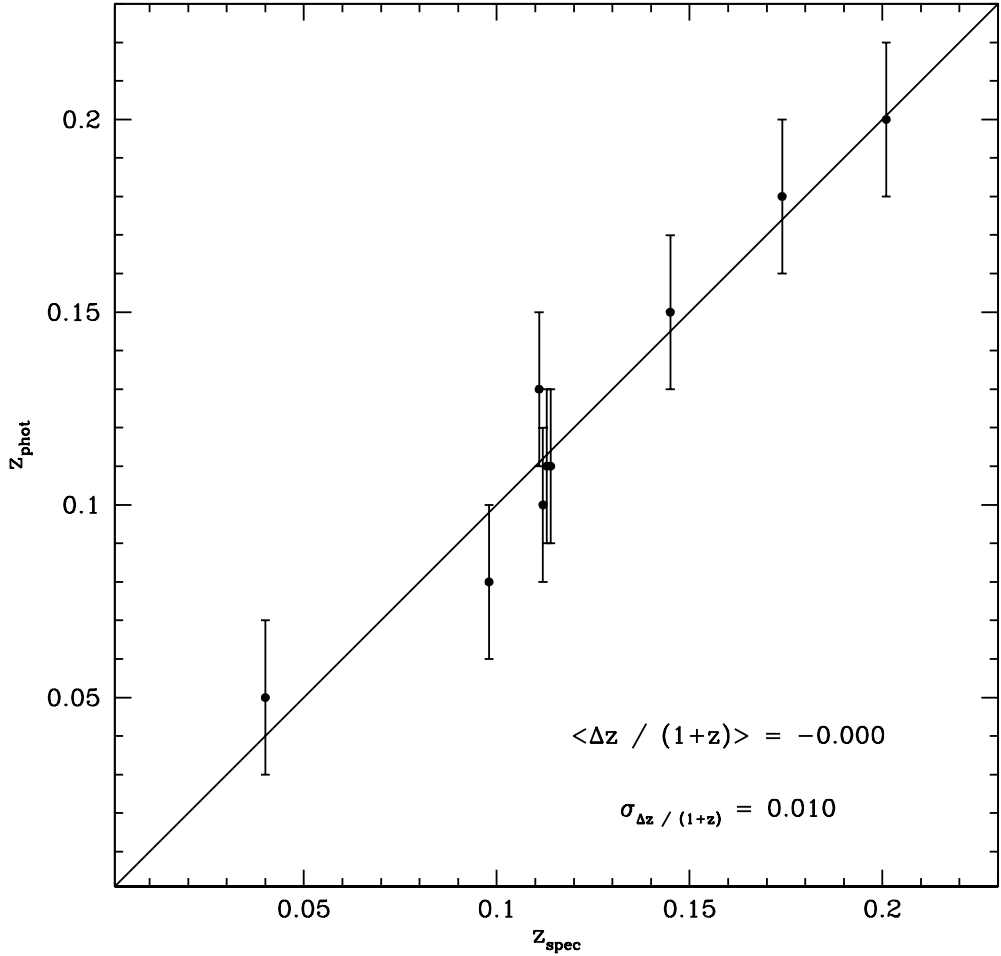}
\caption{Comparison of spectroscopic and photometric redshifts for the
available sample. This plot makes use of the results obtained with the
S\'anchez (2008, {\it in preparation}) code described in the text. The
statistics refer exclusively to the nine points represented in the
plot.}
\label{zphot}
\end{figure*}

The empirically calibrated library of Ben\'\i tez \etal (2004) has
been shown to be precise enough to detect and calibrate photometric
zero points errors within $2-3\%$ as the ones in the NIC3 observations
of the Hubble Ultra Deep Field (see Coe \etal 2006). Using this
calibration technique to correct the COMBO-17 photometry presented in
Hildebrant, Wolf and Ben\'\i tez (2008) reduces the photometric redshift
error from $\delta z/(1+z)$ from $0.038 \pm 0.035$ to $0.001 \pm
0.023$.  We therefore carry out a similar zero point recalibration
using the 9 galaxies with spectroscopic redshifts.  This recalibrated
photometry is then fed to the three codes mentioned above. As
expected, the agreement among them is good for bright objects, but
somewhat breaks down at fainter magnitudes.
 
It is noteworthy that code c) predicts extremely well the redshifts of
the spectroscopic sample: its photometric error is $\delta z \approx
0.000 \pm 0.010$ close to the theoretical accuracy (see
Figure \label{zphot}).  The BPZ software yields $\delta z \approx
0.002 \pm 0.014$, but it mistakes the reddening of an edge-on spiral
at $z=0.09$ for a higher redshift, producing an outlier.

We will further test and refine our photometric redshift techniques,
and the final strategy will probably include a combination of them, in
order to make our measurements as robust as possible. As a token of
the quality of the spectral information, we present the
ALHAMBRA-spectrum of one of the SDSS galaxies in Figure~\ref{spec}
(SDSS coordinates 356.331030,+15.479499, redshift
$z_{sp}=0.113525$). Our best-fit redshift for this object is
$z_{ph}=0.12$. The magnitude offset between the SDSS spectrum and our
photometry is present in the original data, and it must be due to the
smaller flux falling within the SDSS spectrograph fiber. Our data,
even with the preliminary calibration we are using at the moment,
perfectly reproduce even some of the minor details in the spectrum.

\begin{figure*}
\centering
\includegraphics[angle=0,width=14cm]{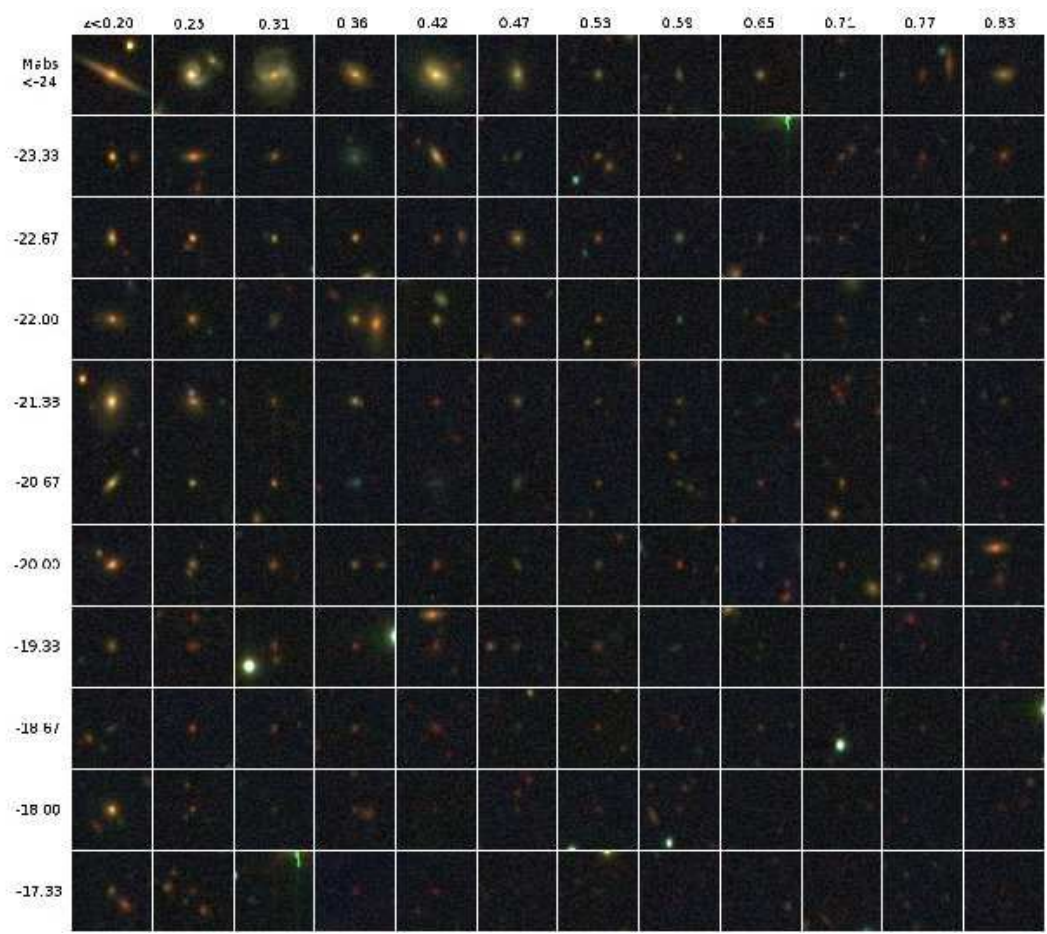}
\caption{Postage stamp images of a set of galaxies extracted from the
  catalog of the field presented in this work. Each postage stamp has
  been generated using the $\lambda$3960\AA~ (blue),
  $\lambda$6750\AA~(green) and Ks-band (red) frames to create a
  three-color image. The galaxies are ordered by increasing redshift
  in the X-axis and decreasing $\lambda$7370\AA~ absolute magnitude
  along the Y-axis, with each plotted galaxy being the most luminous
  of its corresponding redshift-luminosity box. The luminosity ranges
  from M$_{7370} < -$ 24 to M$_{7370} \approx -$ 16.67, and the
  redshift ranges from $z<0.20$ to $z=0.90$. Redshift and magnitude
  values for each column and line are given in the figure.}
\label{poststamp}
\end{figure*}

In Figure~\ref{poststamp} we present a mosaic of 3-color galaxy images
sorted according to redshift (lines) and luminosity (columns). They
span a large range in absolute magnitude (as measured in the band
centered at 7370\AA), from brighter than M(AB) = $-$24, to M(AB)
$\approx -$16.7. The figure illustrates the depth of this preliminary
catalogue and the redshift values that can be obtained.

One of the defining characteristics of any photometric redshift survey
is its effective depth, i.e. the magnitude limit at which it is still
possible to measure precise photometric redshifts. If the photometric
redshifts are estimated using a Bayesian formalism, it is possible to
define a quantity called ``odds'' which serves as powerful quality
indicator of the reliability and accuracy of a photometric redshift
estimation (Ben\'\i tez 2000, Ben\'\i tez \etal 2004). The value of
the odds represents the fraction of the probability $p(z)$
concentrated around the maximum $z_{ph}$. At fainter magnitudes the
photometric noise degrades the redshift information and often $p(z)$
is multimodal or presents a single peak of large width, making
unfeasible an unambiguous estimate of the redshift. Therefore by
imposing cuts on the value of the odds we can select galaxy subsamples
for which the redshift estimates are reliable and accurate; in fact it
can be shown that the redshift accuracy depends on the severity of the
threshold cut.
  
\begin{figure*}
\centering
\includegraphics[angle=0,width=14cm]{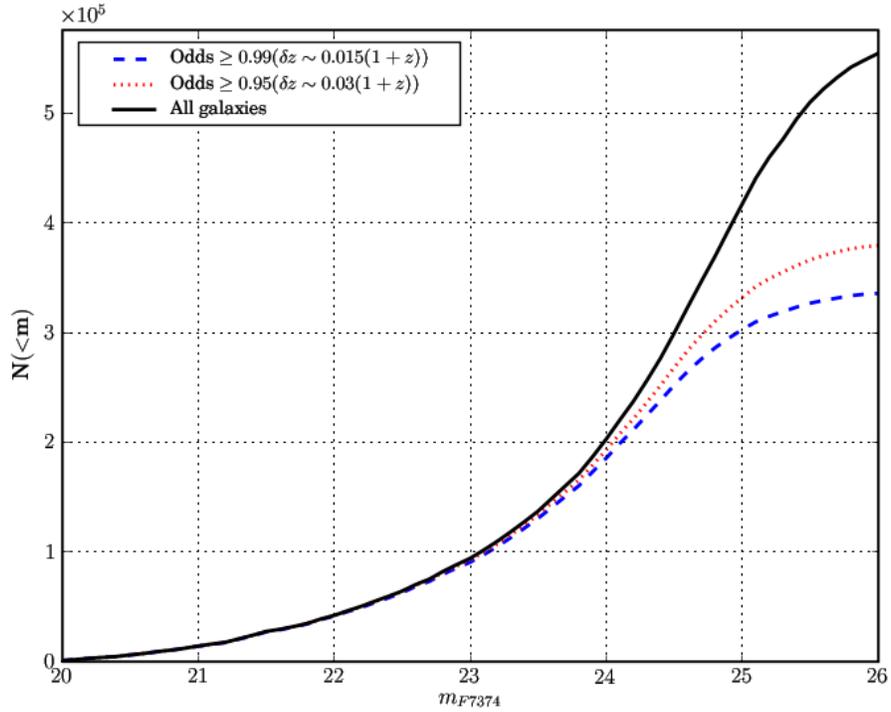}
\caption{ The total number of galaxies we expect to detect in the
  ALHAMBRA-Survey (thick continuous), where we use the $F7374$ band as
  a proxy for the $I$ band.  The dashed and dotted line give the
  total number of galaxies with different odds thresholds (see text).
  These results are based on our first complete $15'\times15'$ arc
  min$^2$ pointing. Our simulations predict the galaxies with odds $>
  0.99$ and odds $>0.95$ will have photometric redshift errors of
  $\delta \lesssim 0.015(1+z)$ and $\delta \lesssim 0.03(1+z)$
  respectively. }
\end{figure*}

\begin{figure*}
\centering
\includegraphics[angle=0,width=14cm]{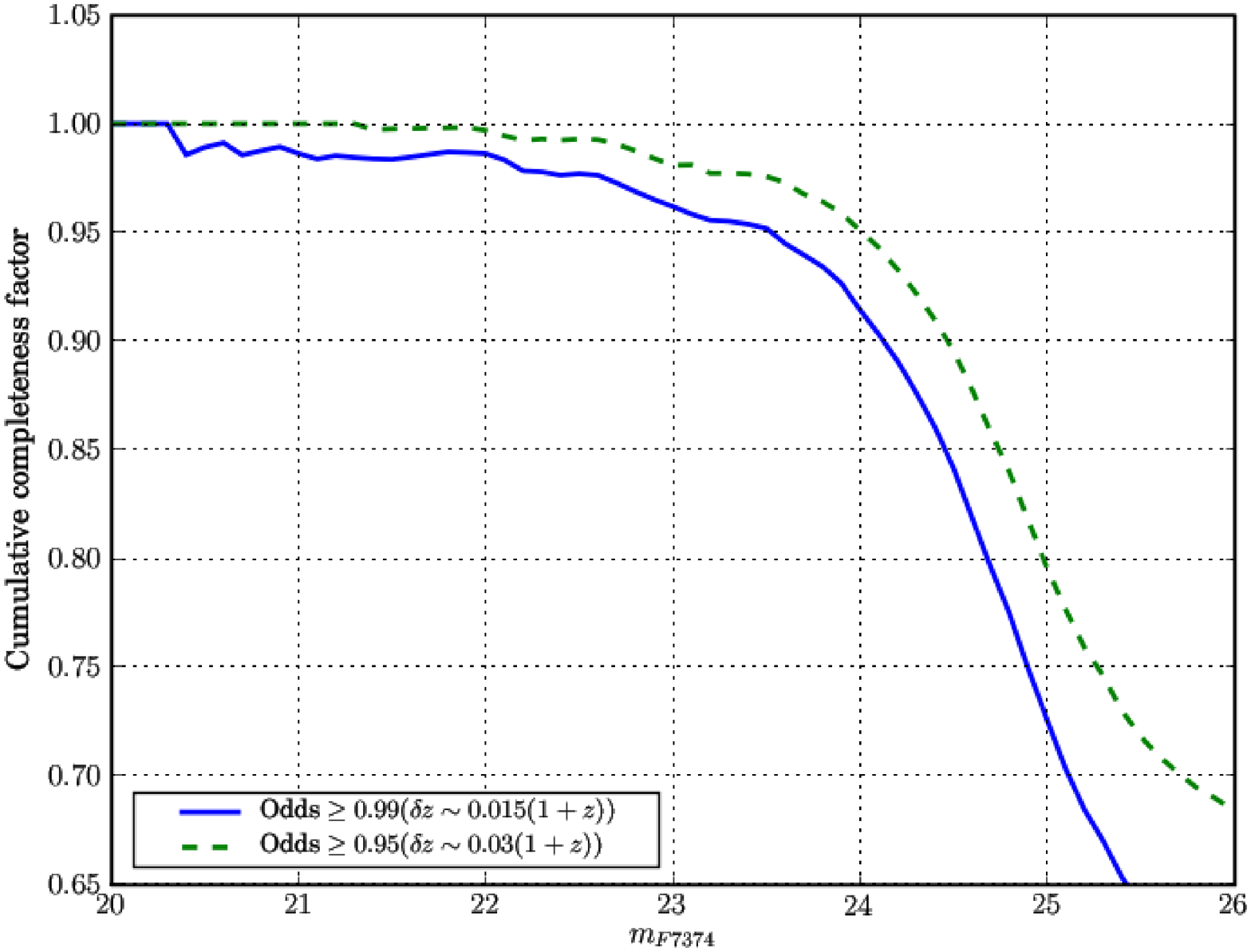}
\caption{ The expected completeness level for the ALHAMBRA-Survey as a
  function of magnitude; here we use the $F7374$ band as a proxy for
  the $I$ band.  The dashed and dotted line correspond to different
  odds thresholds (see text).  These results are based on our first
  complete $15'\times15'$ arcmin$^2$ pointing. Our simulations
  predict the galaxies with odds $> 0.99$ and odds $>0.95$ will have
  photometric redshift errors of $\delta \lesssim 0.015(1+z)$ and
  $\delta \lesssim 0.03(1+z)$ respectively. }
\label{Compl}
\end{figure*}

\section{Comparison with other surveys}

\begin{figure*}
\centering
\includegraphics[angle=-90,width=14cm]{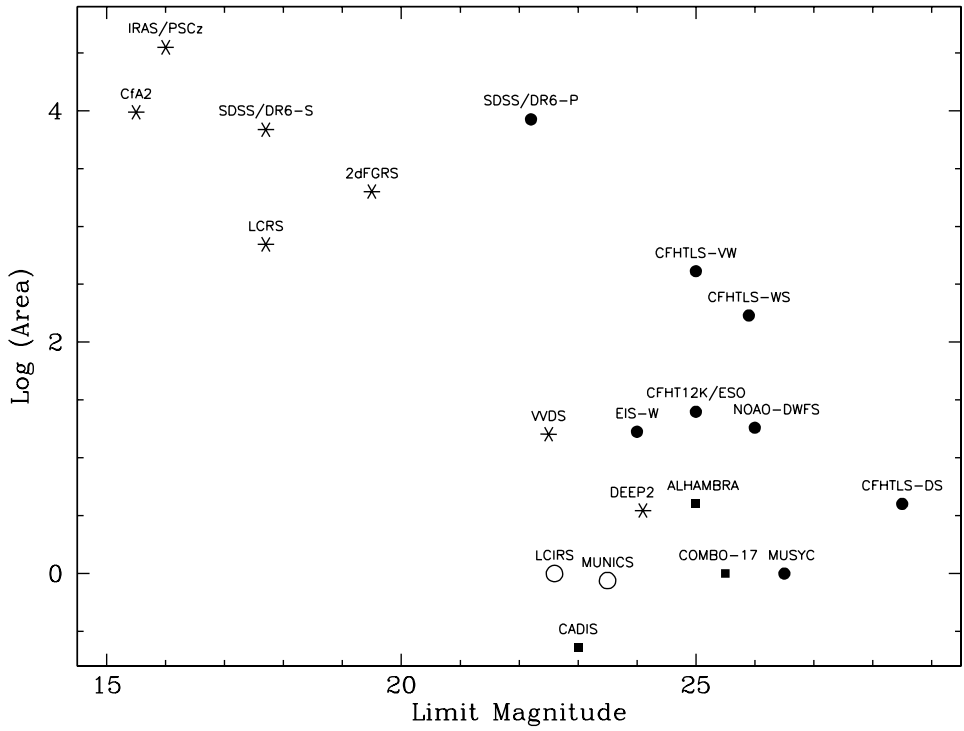}
\caption{The position of different surveys in the area-depth
plane. Only photometric surveys with a minimum of 5 filters and
covering 0.5 square degree at least (with the exception of CADIS) are
shown. Circles correspond to broad band photometric surveys, stars to
spectroscopic surveys, and squares to CADIS, COMBO-17 and
ALHAMBRA. Empty circles correspond to NIR imaging
surveys.}
\label{surveys}
\end{figure*}

Several large, deep photometric and spectroscopic surveys, have been
completed in the last years or are currently being
conducted. Comparison between different surveys, each of them designed
for specific purposes and with different strategies, is not
straightforward. To put the ALHAMBRA-Survey in the context of other
similar efforts we have plotted in Figure~\ref{surveys} their
positions in an area-depth plane\footnote{Notice that the magnitude
limits correspond in general to different spectral bands. In the
particular case of IRAS/PSCz a typical $B$ magnitude has been taken as
indicative of the population, selected in the 60$\mu$ band.}. Only
photometric surveys with 5 or more filters, i.e., those that can
provide general redshift information, and area covered $\geq$ 0.5
square degree (with the exception of CADIS) were included. As we can
see in this plot, the ALHAMBRA-Survey occupies a position in between
the first generation of wide field photometric surveys--amongst which
clearly COMBO-17 has played the leading role up to now--and the
ongoing projects which try to map much larger areas with comparable
depth like Pan-Starrs, DES, PAU, and LSST.  COMBO-17 covers an area
slightly larger than 1 square degree in five fields observed with the
Wide-Field Imager at the MPG/ESO 2.2m telescope in La Silla
(Chile). The large number of bands (five broad band filters $UBVRI$
and 12 medium-band filters from 3500 \AA\ to 9300 \AA) has provided
accurate measurements of photometric redshifts (Wolf \etal 2001a) to
undertake ambitious scientific projects, being two of the last
scientific achievements the measurement of galaxy clustering at
moderate redshifts ($\langle z \rangle =0.6$) (Phleps \etal 2006) and
the analysis of 3D weak lensing (Kitching \etal 2007).

As we have already indicated, the ALHAMBRA survey has been designed to
provide a deeper photometric survey in the Northern hemisphere making
use of 20 optical medium-band filters designed and optimised for the
accurate determination of photometric redshifts (Ben\'\i tez \etal
2008), covering four times the area covered by COMBO-17, and including
from its design the standard NIR filters J, H and K. We are using in
ALHAMBRA the 3.5 m telescope at Calar Alto, and therefore the gain in
aperture and the use of infrared filters will make the ALHAMBRA survey
deeper than COMBO-17. It is worth to say that also the COMBO-17 team
has planned a NIR extension of its survey that will allow to reach $z
\sim 2$ within its covered area (Meisenheimer \etal, MPI technical
report, 2005).

The ALHAMBRA-Survey compares favourably with other deep, large
photometric surveys, even though it uses medium-band filters. It is
only second in limit magnitude to a few broad band photometric
surveys. Indeed, comparing the ALHAMBRA survey (not to mention
spectroscopic surveys) with broad band ones in those terms is unfair,
since the resulting accuracy in redshift and SED determination
precision is significantly different. Thus, whereas typical redshift
accuracies from those broad band surveys is $\sim 0.1$ in $\Delta
z/(1+z)$ at best, we expect to reach 0.015 with ALHAMBRA for several
hundred thousand galaxies.

A recent effort quantitatively similar to some degree to ALHAMBRA is
the one presented by Mobasher \etal (2007), who have
measured photometric redshifts in the COSMOS survey covering an
area of $1.4^\circ \times 1.4^\circ$ containing 367000 galaxies down
to $i\sim25$ using 16 filters and providing accurate redshifts for
faint galaxies up to $z \sim 1.2$. The reliability of their
measurement has been tested comparing with the spectroscopy subsample
zCOSMOS containing 868 normal galaxies with $z<1.2$ (Lilly \etal
2007).

Spectroscopic surveys do of course achieve the highest redshift and
SED precision but they cannot go as deep as photometric surveys, and
not always have a complete spectral coverage, resulting in complex
selection functions. Besides, the detection limit is not homogeneous
along the spectrum, resulting in a lack of completeness which can
prove very difficult to control. In some cases, for example when there
is only one line in the spectrum, the photometric data can be used to
reduce the ambiguity present in the spectroscopic measurements (Lilly
\etal 2007). Given that our survey, in spite of its photometric
nature, is closer in spirit to spectroscopic surveys, we discuss in
the next paragraphs how it compares with them.

\begin{figure*}
\centering
\includegraphics[angle=-90,width=14cm]{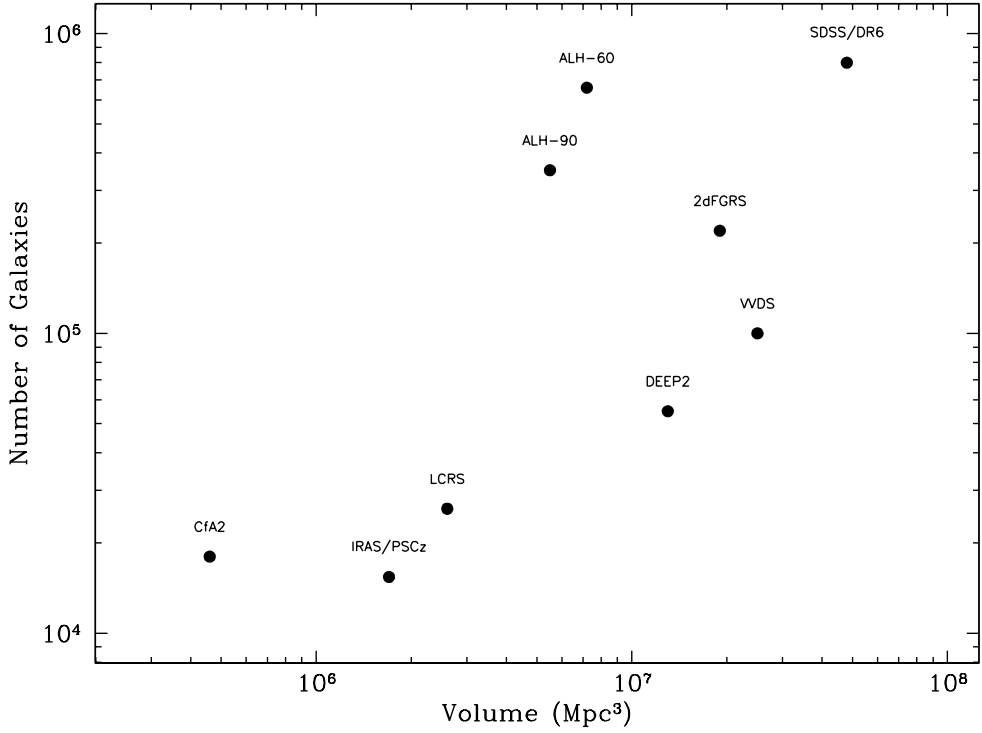}
\caption{Number of detected objects with accurate redshift {\sl
versus} the surveyed volume for spectroscopic surveys covering 0.5
square degrees at least, and the ALHAMBRA-Survey. We have considered a
median redshift of 0.02 for CfA2 and IRAS/PSCz, 0.08 for LCRS, 0.1 for
SDSS/DR6 and 2dF, 0.7 for VVDS, 1.0 for DEEP2, and 0.63 and 0.74 for
ALHAMBRA-90 (complete at the 90\% level) and ALHAMBRA-60 (complete at
the 60\% level) respectively. The plotted ALHAMBRA data are for
$\Delta z/(1+z) \leq 0.03$. }
\label{surveys_s}
\end{figure*}

Indeed, the spectroscopic surveys can provide much more detailed
information about individual objects than any medium-band photometric
survey. However, for all those purposes where that detailed
information is not needed, a survey like ALHAMBRA will prove
advantageous due to the homogeneity of the detection in the different
filters and the ability to produce accurate results even near the
detection limit. Thus, as we wrote before, we expect to be 60\%
complete down to I = 24.7 mag with accurate $z$-determination, with a
median redshift of 0.74. In Figure~\ref{surveys_s}, we have plotted
the expected global performance of different spectroscopic surveys in
terms of surveyed volume and number of objects with good SED and
redshift determinations. The ALHAMBRA-Survey appears close to the SDSS
in terms of number of detected objects, and close to the deepest
surveys in terms of its median redshift.

Another important advantage for the ALHAMBRA-survey is the spectral
range covered. It is in fact the only survey covering the whole
optical domain, from 3500~\AA~ to 9700~\AA, plus the 3 standard NIR
bands. This implies that we are sensitive, within the detection
limits, to any kind of object at any redshift. This not only avoids
the presence of selection functions, but, more important, will
straightforwardly allow the comparative analysis of data at different
redshift values.

We have compiled in Table 2 the relevant data for the spectroscopic
surveys plotted in Figure~\ref{surveys_s} (the number of objects in
the ALHAMBRA survey is given for $\Delta z/(1+z) \leq 0.03$ and, in
parenthesis, $\leq 0.015$). For surveys with a similar (even if
smaller) spectral coverage, like SDSS or 2dFGRS, the ALHAMBRA-Survey
will be 7.8 and 6.7 magnitudes deeper, respectively, what will allow,
in particular, for a detailed analysis of the distant/faint
Universe. At the other extreme, surveys like VVDS or DEEP2 that are
not too far from the ALHAMBRA-Survey in terms of limit magnitude (even
if they are shallower by 1.2 and 2.7 magnitudes, respectively), have
significantly shorter spectral coverage than ALHAMBRA in the optical
domain, and do not include near-IR information.

Hickson \& Mulrooney (1998) presented the first results of the medium
band photometric survey proposed by Hickson \etal (1994). We notice
that the survey used more filters than ALHAMBRA, thus giving a finer
spectral sampling, but their spectral coverage is shorter, from 4450
to 9480\AA. The observing conditions were such that the survey could
only reach m $\approx$20.4, at the 50\% completeness limit. Thus, even
if similar in many aspects to our survey, a real comparison cannot be
made given the bright detection limit they achieved.

In relation with other medium band photometric surveys, covering the
whole spectral range, we have mentioned COMBO-17 and CADIS---which has
also has made use of one of the Calar Alto telescopes---that comprise
a mixed set of broad and narrow band filters. The complete spectral
coverage is assured only by the broad band filters. Their specificity
relies on the use of narrow filters located at some given fixed
positions.  The total area that they cover (1 and 0.2 square degrees
respectively), the photometric depth, and the lack of near-IR
information, are also important differences with the ALHAMBRA-Survey.

Yet another project making use of photometric redshifts and covering
very large areas of the sky, but with depth less than $z \sim 0.7$, is
MegaZ-LRG (Collister \et 2007), containing about a million SDSS
luminous red galaxies, lying in a region of 5915 square degrees, with
limiting magnitude $i<20$. This catalog, with a surveyed volume of
about 2.5 h$^{-3}$ Gpc$^3$, has been recently used to measure
cosmological parameters from the large scale structure in a way
competitive with the shallow wide-angle spectroscopic surveys (Blake
\etal 2007). We must remark that an $M_{\star}$ galaxy lying at
redshift $z\sim 1.4$ will be included in ALHAMBRA, since its visual
magnitude should be about 24. At this redshift, the whole ALHAMBRA
survey covering 4 square degrees will produce a volume of about $3
\times 10^7 \,h^{-3}$Mpc$^3$.

Other ongoing projects are the ESO/VST KIDS that will cover 1400
square degrees making use of images in 4 broad bands. Its major goal
will be the study of weak lensing, although it should also be a good
sample to study baryonic acoustic oscillations (Peacock \et 2006), and
the Dark Energy Survey, that makes use of a new camera on the CTIO 4m
telescope. This survey will cover 5000 square degrees, but no near-IR
observations are planned. Farther into the future, the LSST plans to
map about half of the sky in the standard broad bands $ugrizy$ to
obtain about $3\times10^9$ photometric redshifts. The first light of
this project is scheduled for 2012, and as Peacock \et (2006) have
pointed out, the lack of near-IR information may be an issue.

Different research projects devoted to the study of the evolution of
galaxies across the Hubble time will get great advantage from the
ALHAMBRA survey, when the complete catalog is released as publicly
available. For example, this survey will increase the reliability of
the present studies of the growth of the population of red galaxies
since $z \sim 1$ (Ford \et 2006) or the evolution with redshift of the
color-density relation (Cucciati \et 2006, Cooper \et 2007). Other
field where a deep and wide catalog like ALHAMBRA will be very useful
is the analysis of the assembly history of red galaxies (Brown \et
2008) or field spheroidals (Treu \et 2005). Similarly, the ALHAMBRA
survey covers enough area and is deep enough to trace accurately the
evolution of the luminosity function for different spectral types
(Poli \et 2003, Treu \et 2005, Wolf \et 2003, Zucca \et 2006,
Marchesini \et 2007) and the growth of stellar mass with cosmic time
(Borch \et 2006).

All in all, we can say that even if ALHAMBRA is a photometric survey,
it shares important aspects with the spectroscopic surveys, with which
it compares well in many respects. Its advantages, depending on the
proposed goals, are the complete spectral coverage, the homogeneity in
the detection level along the spectrum, and the depth.

\begin{table}
\caption{Main characteristics of wide field ($\geq 0.5$ square
degrees) spectroscopic surveys} \centering
\label{Surveys}
\begin{tabular}{lcccc}
\hline
Survey  &  Area $^{\square}$ & Spectral range (\AA)& $z$ (median) & N$_{objects}$ \\
\hline
CfA/SRSS & 18000 & 4300-6900 &  0.02 & 1.8 $\times$ 10$^4$\\
SDSS/DR6  &  6860 & 3800-9200 & 0.1 & 7.9 $\times$ 10$^5$ \\
LCRS      &  700  & 3350-6750 & 0.1 & 2.6 $\times$ 10$^4$\\
2dFGRS    &  2000 & 3700-8000 & 0.11 & 2.2 $\times$ 10$^5$ \\
VVDS    & 16 & 5500-9500 & 0.7 &  1.0 $\times$  10$^5$ \\
DEEP2  & 3.5  & 6500-9100 & 1.0 & 5.5 $\times$ 10$^3$  \\
\hline
ALHAMBRA-60 & 4 & 3500-9700 ($+JHK$) & 0.74 & 6.6 (3.0) $\times$ 10$^5$ \\
ALHAMBRA-90 & 4 & 3500-9700 ($+JHK$) & 0.63 & 3.5 (1.0) $\times$ 10$^5$ \\
\hline
\end{tabular}
\end{table}
%

\section{Final Considerations}

The ALHAMBRA-Survey places itself halfway between relatively shallow,
limited spectral coverage, wide-area spectroscopic surveys, and deep,
large area, broad band photometric surveys. Trying to optimally
combine the advantages of each kind of survey, we intend to observe a
large area (4 $^{\square}$) using a specially designed set of 20
medium band, minimally overlapping filters covering the whole visible
range from 3500 \AA\ to 9700 \AA, plus the standard $JHK_s$ near
infrared filters. It will provide homogeneous data down to AB
$\approx$ 25 for all the filters from 3500\AA~ to 8500\AA, with a
magnitude limit of AB = 23.2 at 9550 \AA. The filter characteristics
where decided to allow us to detect even relatively faint emission
features (observed EW $\geq$ 30\AA).

Together with the NIR information to $K_s \approx$ 20, $H \approx$
21, $J \approx$ 22 (in the Vega system), this will allow the
measurement of the redshift and SED for several hundred thousand
objects. Indeed, the survey was designed having in mind the use of
photometric redshift techniques as the basic analysis tool. We have
carried out detailed simulations based on available deep
catalogs. We expect that the ALHAMBRA-Survey will be able to produce
high-quality redshifts and accurate spectral types for more than
600,000 galaxies down to $I_{AB} \approx 24.7$, with redshift accuracy
$\Delta z/(1+z)\approx0.015$. The first data and preliminary results
we present here confirm that these expectations are realistic.

With its volume surveyed, median redshift and number of objects with
accurate redshift and SED determination, the ALHAMBRA-Survey will
provide a unique set of data for many different studies in different
astrophysical and cosmological domains.

  The main objective of our survey is the study of cosmological
evolution, under the many facets it can offer. We will study the
evolution of the large scale structure, the number and content of
clusters at different redshifts, the evolution of the populations of
different cosmic objects, and the processes leading to galaxy
formation, evolution, and differentiation. The unbiased nature of the
survey will also allow for the study of many different kinds of
objects, ranging from emission-line galaxies to the diverse types of
AGNs, and stars in our own Galaxy.

\begin{acknowledgements}

We acknowledge the decisive support given by the ALHAMBRA Extended
Team to the project (see \url{http://www.iaa.es/alhambra} for the
details regarding the project implementation and organization). We
also wish to acknowledge the Calar Alto staff for their warm
assistance for a fruitful start of the observations. We acknowledge
the work of C. C\'ardenas to measure the filter characteristics and of
V. Peris for his work to produce the combined pictures out of the
ALHAMBRA images. We thank our anonymous referee, whose comments helped
to improve the clarity and quality of our manuscript. The authors
acknowledge support from the Spanish Ministerio de Educaci\'on y
Ciencia through grants AYA2002-12685-E, AYA2003-08729-C02-01,
AYA2003-0128, AYA2004-08260-C03-01, AYA2004-20014-E, AYA2004-02703,
AYA2004-05395, AYA2005-06816, AYA2005-07789, AYA2006-14056, and from
the {\sl Junta de Andaluc\'{\i}a}, TIC114, TIC101 and {\sl Proyecto de
Excelencia} FQM-1392. NB, JALA, MC, and AFS acknowledge support from
the MEC {\sl Ram\'on y Cajal} Programme. NB acknowledges support from
the EU IRG-017288.  This work has made use of software designed at
TERAPIX and the Canadian Astronomy Data Centre.

\end{acknowledgements}
\end{document}